\documentclass[12pt]{article}
\pdfoutput=1
\usepackage{natbib}
\usepackage{multirow}
\usepackage{amsthm}
\usepackage{amssymb}
\usepackage{xcolor}
\usepackage{bbm}
\usepackage{lscape}
\usepackage{caption}
\usepackage{subcaption}
\usepackage{mathtools}
\usepackage{url}
\usepackage{float}
\usepackage{amsmath}
\usepackage{mathalfa}

\usepackage{algorithm}
\usepackage[parfill]{parskip}
\usepackage{algorithmic}
\usepackage[T1]{fontenc}
\usepackage{titling}
\usepackage[toc,page]{appendix}
\usepackage{listings}
\usepackage{geometry}
\usepackage{dsfont}
\usepackage{wrapfig}
\usepackage{graphicx}
\usepackage[english]{babel}
\newtheorem{theorem}{Theorem}[section]
\newtheorem{corollary}{Corollary}[theorem]
\numberwithin{equation}{section}

\newcommand{\beginsupplement}{%
        \setcounter{table}{0}
        \renewcommand{\thetable}{\arabic{table}}%
        \setcounter{figure}{0}
        \renewcommand{\thefigure}{\arabic{figure}}%
        \setcounter{section}{0}
        \renewcommand{\thesection}{Web Appendix \Alph{section}}%

\captionsetup[figure]{labelfont={default},labelformat={default},labelsep=period,name={Web Figure}}

\captionsetup[table]{labelfont={default},labelformat={default},labelsep=period,name={Web Table}}

\captionsetup[section]{labelfont={default},labelformat={default},labelsep=period,name={Web Appendix}}

     }

\definecolor{ao}{rgb}{0.0, 0.5, 0.0}

\title{A multi-arm multi-stage platform design that allows pre-planned addition of arms while still controlling the family-wise error }
\author{Peter Greenstreet, Thomas Jaki, Alun Bedding, Chris Harbron, Pavel Mozgunov}
\date{}
\begin{document}
\maketitle

\begin{abstract}
There is growing interest in platform trials that allow for adding of new treatment arms as the trial progresses as well as being able to stop treatments part way through the trial for either lack of benefit/futility or for superiority. In some situations, platform trials need to guarantee that error rates are controlled. This paper presents a multi-stage design that allows additional arms to be added in a platform trial in a pre-planned fashion, while still controlling the family wise error rate. 
A method is given to compute the sample size required to achieve a desired level of power and we show how the distribution of the sample size and the expected sample size can be found. A motivating trial is presented which focuses on two settings, with the first being a set number of stages per active treatment arm and the second being a set total number of stages, with treatments that are added later getting fewer stages. Through this example we show that the proposed  method results in a smaller sample size while still controlling the errors compared to running multiple separate trials. 
\end{abstract}

%




%

\section{Introduction}
\label{sec:introduction}
Clinical trials take many years to run and during this time it is not uncommon for new promising treatments to emerge that warrant evaluation. It may be advantageous to include these treatments into an ongoing trial, due to the shared trial infrastructure and the possibility to use a shared control group. This can result in useful therapies being identified faster while reducing 
cost and time~\citep{cohen2015adding}. The trial potentially requires less administrative and logistical effort than setting up separate trials, so can noticeably speed up the development process \citep{BurnettThomas2020Aeta}. The addition of more arms may also enhance the recruitment, as patients have a higher chance of receiving an experimental treatment, therefore, making them potentially more likely to join a trial \citep{MeurerWilliamJ2012ACTA}.
\par
There is an ongoing discussion about how to add new treatments to clinical trials \citep{lee2021}.
\cite{cohen2015adding} conclude in their literature review that there is no systematic or comprehensive method for adding new arms to clinical studies. Recently, several approaches to adding treatment arms have been proposed which aim to help tackle this issue. \cite{MaxineBennett2020Dfaa} and \cite{Choodari-OskooeiBabak2020Anea} propose approaches which extend the Dunnett test \citep{DunnettCharlesW1955AMCP} to allow for unplanned  additional arms to be included into multi-arm trials while still controlling the family wise error rates (FWER). This methodology does not incorporate the possibility of interim analyses. 
\par
Interim analyses are a further way to potentially improve the design of a clinical trial \citep{WasonJames2016Srfm}.
They allow for ineffective treatments to be dropped for futility (or lack of benefit) earlier, as well as allowing the trial to stop early if a superior treatment is found. Both of these can result in the reduction of the expected sample sizes and costs of a trial.  Multi-arm multi-stage (MAMS) design  \citep[e.g.][]{MagirrD.2012AgDt} allows for several treatments to be evaluated within one study and incorporate interim analyses for efficiency, but does not allow for  additional arms to be added throughout the trial. \cite{BurnettThomas2020Aeta} developed an approach that  incorporates unplanned additional treatment arms to be added to a trial already in progress using the conditional error principle \citep{ProschanMA1995DEoS}, to allow for modifications during the course of a trial. The unplanned nature of the adaptation, however, means that type I error and power for different arms may be different. As a result,  the additional treatments can be underpowered.
\par
In this work, we provide an analytical method for adding of treatments to a multi-arm multi-stage (MAMS)  trial in a pre-planned manner, while still controlling the statistical errors. The design assumes that, at the design stage, it is known that new treatments will be added later into the trial. Unlike the currently developed approaches, the approach proposed allows for a trial with multiple stages and multiple arms to be designed, so that pre-planned additional treatments can be added to the trial while still controlling the family wise error rate (FWER) in the strong sense \citep{dmitrienko2009} and achieve suitable power for the trial. 
\par
We focus our investigation on two settings: (i) each active treatment has the same number of stages; (ii) there is a fixed total number of stages. Using these two settings, we revisit the design of FLAIR \citep{HowardDenaR2021Apti} that had a treatment arm added during the trial. We derive the critical boundaries, the sample sizes and the expected sample sizes for trials based on FLAIR, and study the effect on errors when deviating from the planned additions. These two settings are compared to running separate trials, using the MAMS design by \cite{MagirrD.2012AgDt} and using the original trial design without adjusting for additional treatments.
\section{Method}
\label{sec:Method}
\par
\subsection{Setting}
Consider a clinical trial with up to $K$ experimental arms that will be tested against one common control arm with $K^\star$ experimental arms starting at the beginning of the trial, where $K^\star \geq 1$, and $K-K^\star$ arms being added later. The primary outcome on each patient is independent and normally distributed with known variance $\sigma^2$. In total, the control treatment is recruited for a maximum of $J_0$ stages with there being a maximum of $J_0-1$ interim analyses, with an analysis taking place at the end of each stage. Each of the active treatments can have any number of stages (provided it is pre-specified and their total number is less or equal to $J_0-1$) which coincide with the analysis for the other treatments. Each additional treatment can be added at any of the interim analyses as long as this is pre-planned at the design stage of the trial development. When comparing the control to the active treatments only the concurrent controls are used in the comparisons. This means only participants recruited to the control arm at the same time as the active arm are used in the comparisons.
\par
The null hypotheses of interest are
$
H_{01}: \mu_1 \leq \mu_0, H_{02}: \mu_2 \leq \mu_0, ... , H_{0K}: \mu_K \leq \mu_0,
$
where $\mu_1, \hdots, \mu_K$ are the mean responses on the $K$ experimental treatments and $\mu_0$ is the mean response of the control group. The global null hypothesis, $\mu_0=\mu_1=\mu_2=\hdots=\mu_K$, is denoted by $H_{G}$. Each of the $K$ hypotheses is potentially tested at a series of analyses indexed by $j=1,\hdots,J_k$ where $J_k$ is the maximum number of analyses for a given treatment $k = 1,\hdots, K$. Let $J$ denote the maximum number of planned analyses of any of the active treatments, $J=\max_k(J_k)$. The total number of stages of the control treatment is $J_0$. Let $s(k)$ be the stage when treatment $k$ is added to the trial  and define the vector of adding times by $S=(s(1),\hdots,s(K))$. We denote the ratio of patients recruited to treatment $k$ by the end of its $j^\text{th}$ stage by $r_{k,j}$ and denote $n_{k,j}$ as the  number of patients recruited to treatment $k$ by the end of its $j^\text{th}$ stage with $k=0$ denoting the control treatment. The number of patients recruited to the first stage of treatment $k$ is defined as $n_k$ so that $n_{k,j}=n_k \frac{r_{k,j}}{r_{k,1}}$. The total sample size of a trial is denoted by $N$, where the maximum total planned sample size, $\max(N)= \sum_{k=0}^K n_{k,J_k}$. At analysis $j$ for treatment $k$, to test $H_{0k}$ it is assumed that responses, $X_{k,i}$, from patients $i=1,\hdots, n_{k,j} $ are observed, as well as the responses $X_{0,i}$ from patients $i=n_{0,s(k)}+1,\hdots, n_{0,s(k)+j} $, which are the outcome of the patients allocated to the control which have been recruited since treatment $k$ has been added into the trial up to the $j$th analysis of treatment $k$.
The test statistics
\begin{equation*}
Z_{k,j}=\frac{ n_{k,j}^{-1} \sum^{n_{k,j}}_{i=1} X_{k,i} - (n_{0,s(k)+j}-n_{0,s(k)})^{-1} \sum^{n_{0,s(k)+j}}_{i=n_{0,s(k)}+1} X_{0,i} }{\sigma\sqrt{(n_{k,j})^{-1}+ (n_{0,s(k)+j}-n_{0,s(k)})^{-1} }},
\end{equation*}
\par
are used to test hypothesis $H_{0k}$. Upper and lower stopping boundaries,  $U_{k}=(u_{k,1},\hdots,u_{k,J_k})$ and $L_{k}=(l_{k,1},\hdots,l_{k,J_k})$, are used for the decision-making as follows. If $ Z_{k,j}> u_{k,j}$ then $H_{0k}$ is rejected and the trial stops with the conclusion that treatment $k$ is superior to control. If $ Z_{k,j}< l_{k,j}$ then treatment $k$ is dropped from all subsequent stages of the trial. If the $Z$ statistics for all the treatments fall below their lower boundary, the trial stops for futility. Treatment $k$ and control continues to its next stage $j+1$ if neither of these conditions are met, so $l_{k,j} \leq Z_{k,j} \leq u_{k,j}$. The boundaries are found to control the family wise error rate (FWER) in the strong sense at a specified desired level $\alpha$  which is defined as
\begin{align*}
P(\text{reject at least one true } H_{0k} \text{ under any null configuation}, k=1,\hdots, K) \leq \alpha.
\label{eq:FWER}
\end{align*}
\par
\begin{figure}[]
\begin{subfigure}{.49\textwidth}
  \centering
  \includegraphics[width=.95\linewidth,trim= 0 2cm 0 0.5cm, clip]{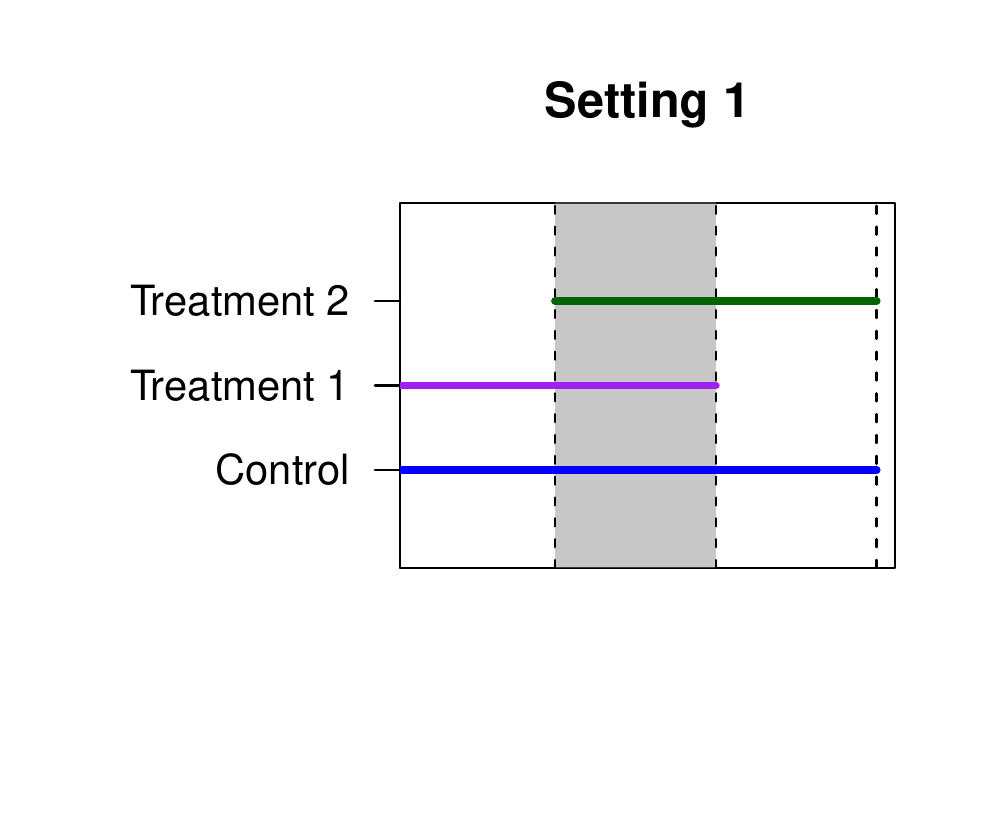}  
\end{subfigure}
\begin{subfigure}{.49\textwidth}
  \centering
  \includegraphics[width=.95\linewidth,trim= 0 2cm 0 0.5cm, clip]{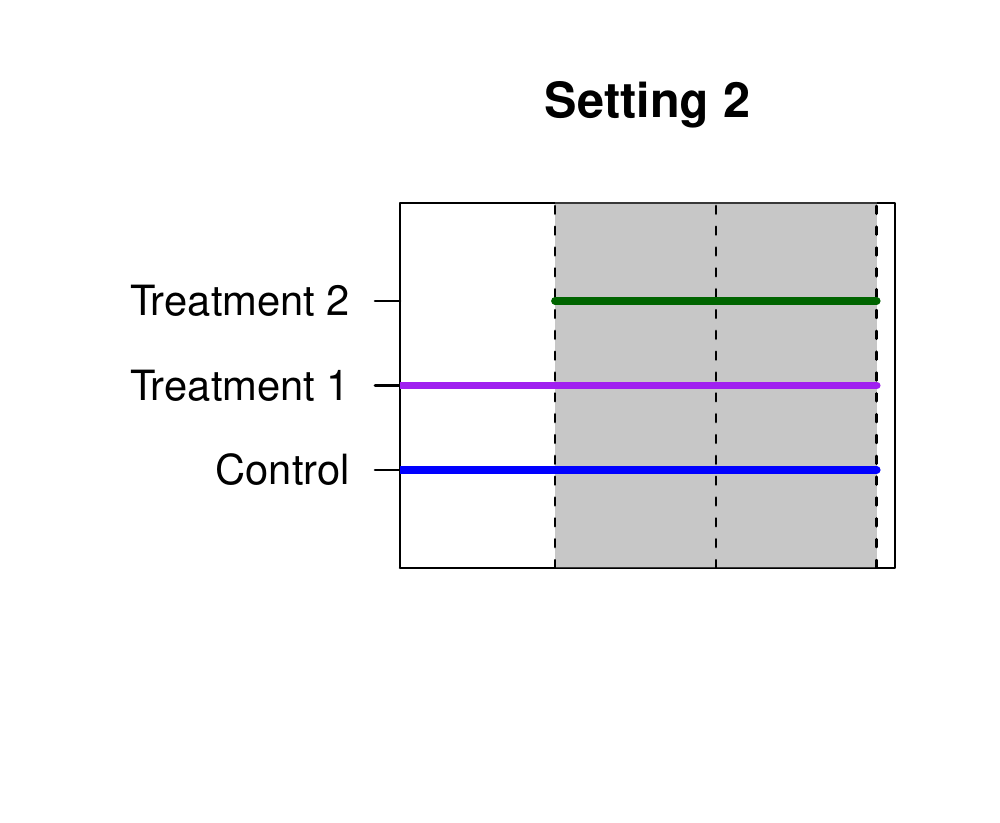}  
\end{subfigure}
\caption{Examples of the two settings for a two active arm setting with $J_0=3$ and Treatment 2 being added at stage 1. For Setting 1, both active treatments get 2 interim analyses, and for Setting 2, Treatment 2 is added one stage and, hence, has one fewer stage. The grey represents areas of possible shared control group. The dashed black line represents an interim analysis.}
\label{fig:screp}
\end{figure}

While this work concerns the general procedure of adding, we focus on two special cases - see Figure \ref{fig:screp}. Setting 1 is the case that each active treatment is planned to have the same number of stages regardless of when it is added. Setting 2 is the case with a set total number of stages with the later a treatment is added the fewer stages are planned for it. Note in Setting 1, $J_1=\hdots = J_K = J$ and $J_0=\max(S)+J$ and in Setting 2, $J_0=J$ and  $J_k= J-s(k)$.
\subsection{Strong control of FWER}
\label{FWERsection} 
Following the method of \cite{DunnettCharlesW1955AMCP}, one can exploit the correlation between the test statistics arising from the common control responses. This description follows \cite{MagirrD.2012AgDt}. For any vector of constants $\Theta=(\theta_1, \hdots, \theta_K)$ and $k=1,\hdots,K,$  $j=1,\hdots, J_k$, letting $
I_{k,j}=\sigma^2( n_{k,j}^{-1}+(n_{0,j+s(k)}-n_{0,s(k)})^{-1})
$, define the events,
\begin{align*}
A_{k,j}(\theta_k)=&[Z_{k,j}<  l_{k,j} + (\mu_k-\mu_0 -\theta_k)I^{1/2}_{k,j}],
\\
B_{k,j}(\theta_k)=&[ l_{k,j}  + (\mu_k-\mu_0 -\theta_k)I^{1/2}_{k,j} < Z_{k,j}  <  u_{k,j} + (\mu_k-\mu_0 -\theta_k)I^{1/2}_{k,j}].
\end{align*}
If $\mu_k -\mu_0=\theta_k$ for $k=1,\hdots, K$, the event that $H_1,\hdots, H_K$ all fail to be rejected is equivalent to
\begin{align*}
\bar{R}_K(\Theta)
& = \bigcap_{k \in \{m_1, \hdots, m_K \} } \Bigg{(} \bigcup^{J_k}_{j=1} \Bigg{[} \bigg{(} \bigcap^{j-1}_{i=1}B_{k,i}(\theta_k) \bigg{)} \cap A_{k,j}(\theta_k) \Bigg{]} \Bigg{)},
\end{align*}
with the convention that $\bigcap^{0}_{i=1}= \Omega$ where $\Omega$ is the whole sample space, $m_1 \in \{1,\hdots, K\} $ and $m_k \in \{1,\hdots, K\} \backslash \{ m_1,\hdots, m_{k-1} \}$. The notation $m_k$ is used to reflect the fact that the order in which treatments are added affects the FWER. 
The events  $A_{k,j}(\theta_k)$ and $B_{k,i}(\theta_k)$ can be rearranged so the following holds
\begin{equation}
P(\bar{R}_K(\Theta))=\int_{-\infty}^{\infty} \hdots \int_{-\infty}^{\infty} \Bigg{[} \prod_{k=1}^K \Bigg{(} \sum_{j=1}^{J_k} \Phi_j(L_{k,j}(\theta_k),U_{k,j}(\theta_k) ,\Sigma_{k,j}) \Bigg{)} \Bigg{]}  d\Phi(t_1) \hdots d\Phi(t_{J_0}),
\label{FWERequation}
\end{equation}
where
\begin{equation}
t_j= \frac{\sum_{i=n_{0,(j-1)}+1}^{n_{0,j}}(X_{0,i}-\mu_0)}{\sigma \sqrt{n_{0,j}-n_{0,(j-1)}}}.
\label{equ:t}
\end{equation}
Here $\Phi(\cdot)$ denotes the standard normal distribution function, and $\Phi_j(L_{k,j}(\theta_k),U_{k,j}(\theta_k),\Sigma_{k,j})$ denotes the result of integrating  the $j$-dimensional normal density with mean zero and correlation matrix, $\Sigma_{k,j}$ with the $(i,i^\star)$th element $(i \leq i^\star)$ of $\Sigma_{k,j}$ is $\sqrt{\frac{r_{k,i} }{r_{k,i^\star}}}$. The integration is over the region defined by a vector of lower limits $L_{k,j}(\theta_k)=(l_{k,1}(\theta_k),\hdots l_{k,j-1}(\theta_k), - \infty)$, and upper limits, $U_{k,j}(\theta_k)=(u_{k,1}(\theta_k),\hdots u_{k,j-1}(\theta_k), l_{k,j}(\theta_k))$, where
\begin{align*}
l_{k,j}(\theta_k)= & l_{k,j} \sqrt{1+\frac{r_{k,j}}{r_{0,s(k)+j}-r_{0,s(k)}}}   +  \frac{\sqrt{r_{k,j}}}{r_{0,s(k)+j}-r_{0,s(k)}} 
\\ & \bigg{(}\sum^j_{i=1} t_{s(k)+i} \sqrt{r_{0,s(k)+i}-r_{0,s(k)+(i-1)}} \bigg{)}  - \frac{\theta_k\sqrt{n_{k,j}}}{\sigma},
\\
u_{k,j}(\theta_k)= & u_{k,j} \sqrt{1+\frac{r_{k,j}}{r_{0,s(k)+j}-r_{0,s(k)}}}   +  \frac{\sqrt{r_{k,j}}}{r_{0,s(k)+j}-r_{0,s(k)}} 
\\ & \bigg{(}\sum^j_{i=1} t_{s(k)+i} \sqrt{r_{0,s(k)+i}-r_{0,s(k)+(i-1)}} \bigg{)}  - \frac{\theta_k \sqrt{n_{k,j}}}{\sigma}.
\end{align*}
\par
Note that, in contrast to \cite{MagirrD.2012AgDt}, the proposed approach accounts for the fact that treatments can be added at different points and hence $l_{k,j}(\theta_k)$ and $u_{k,j}(\theta_k)$  depend on $s(k)$. It also allows for different stopping boundaries per treatment: $A_{k,j}(\theta_k)$ and $B_{k,j}(\theta_k)$ depend on $l_{k,j}$ and $u_{k,j}$ and there are different maximum numbers of stages per treatment. Then, one can obtain the following result.
\begin{theorem}
For any $\Theta$, under the conditions above, $P( \text{reject at least one true } H_{0k} | \Theta) \leq P(\text{reject at least one true } H_{0k} | H_{G})$.
\label{the:FWER}
\end{theorem}

\par
The proof of Theorem \ref{the:FWER} is given in the  Supporting Information in Web Appendix A. It follows from Theorem \ref{the:FWER} that the FWER is maximized under the global null hypothesis.
\par
\begin{corollary} 
Setting $\Theta=\mathbf{0}$ and finding $P(\bar{R}_K(\Theta))$ such that  $P(\bar{R}_K(\Theta))=1-\alpha$ controls FWER in the strong sense at level $\alpha$.
\label{cor:Null}
\end{corollary}
\par
\begin{proof}
Under the global null hypothesis $\mu_0=\mu_k$ for all $k \in 1,\hdots K$ so that $\Theta=(0, \hdots ,0)=\mathbf{0}$. Using Theorem \ref{the:FWER} FWER is controlled in the strong sense at level $\alpha$ if $P(\bar{R}_K(\mathbf{0}))=1-\alpha$.
\end{proof} 
\par 
As a result of Corollary \ref{cor:Null}, the stopping boundaries under the global null hypothesis, which result in $P(\bar{R}_K(\mathbf{0}))=1-\alpha$, will guarantee strong control of FWER at level $\alpha$.
\par
As mentioned above, the proposed methodology allows for different critical boundaries to be used for each treatment $k$ as seen in Equation \eqref{FWERequation}. To find the boundaries one can use the functions $L_{k}=f_k(a_k)$ and $U_{k}=g_k(a_k)$ to reduce the number of unknowns, where $f_k$ and $g_k$ are the functions for the shape of the upper and lower boundaries respectively and $a_k$ are scalar parameters specific to each active treatment. One can use a single parameter $a$  to find the boundaries so $f_k=f_{k'}$, $g_k=g_{k'}$ and $a_k=a_{k'}$ which is similar to the method presented in \cite{MagirrD.2012AgDt}, with the advantage of there being an equal number of unknowns to equations. However, using the same boundaries for each treatment arm, regardless of when it was added, can result in different probabilities of dropping each treatment which might be undesirable. It may be of interest in having a  different stopping boundary shapes for each treatment, as the same shape for each treatment may not be optimal as the trial may need greater sample size compared to a design with different stopping boundary shapes as seen in the Supporting Information (Web Appendix D). This requires using $L_{k}=f_k(a_k)$ and $U_{k}=g_k(a_k)$ which results in $K$ scaler parameters to be found, $\mathbf{a}=(a_1,\hdots,a_K)$.
\par
To calculate $a_k$ for all $k = 1,\hdots,K$, we introduce the requirements on the pairwise error rate (PWER) being the same for all active treatments, where PWER is the probability of rejecting the null hypothesis $H_{0.k}$ incorrectly.  For the PWER for treatment $k$, it is assumed that no active treatment other than $k$ can stop the trial. This assumption eliminates the possibility of stopping the trial without finishing testing treatment $k$, hence, maximises the probability of the corresponding error. The PWER denoted by $\alpha^\star_k$ for treatment $k$ is
\begin{equation}
\alpha^\star_k=1-\sum^{J_k}_{j=1} \Phi(U^\star_{k,j},L^\star_{k,j},\ddot{\Sigma}_{k,j}), 
\label{PWER}
\end{equation}
with $L^\star_{k,j}=(l_{k,1},\hdots, l_{k,j-1}, -\infty)$, $U^\star_{k,j}=(u_{k,1},\hdots, u_{k,j-1}, l_{k,j})$,  and covariance matrix $\ddot{\Sigma}_{k,j}$. The $(i,i^\star)$th element $(i \leq i^\star)$ of $\ddot{\Sigma}_{k,j}$  is
\begin{align*}
\Bigg{(}\sqrt{r_{k,i}^{-1}+(r_{0,s(k)+i}-r_{0,s(k)})^{-1}}\sqrt{r_{k,i^\star}^{-1}+(r_{0,s(k)+i^\star}-r_{0,s(k)})^{-1}} \Bigg{)}^{-1}  \Bigg{(} \frac{1}{r_{k,i^\star}} + \frac{1}{r_{0,s(k)+i^\star}-r_{0,s(k)}}  \Bigg{)}.
\end{align*}
To ensure equal PWER across all the treatments and ensure FWER is controlled the iterative approach in Algorithm \ref{Alg:FWER} is proposed. This approach yields the desired properties as with each iteration we update every $a_k$ so that PWER is equal for all the active treatments and then using step H we ensure that the FWER is controlled by using Corollary \ref{cor:Null}.
\begin{algorithm}[]
\caption{Iterative approach to compute the stopping boundaries}
\begin{itemize}
\item[0] Begin by assuming $\mathbf{a}=(a_1,a_1,\hdots,a_1)$ and find $a_1$ such that $\mathbf{a}$ controls FWER at a specified level, $\alpha$, using Equation \eqref{FWERequation} with $\Theta=0$. Then repeat the following iterative steps until each element of $\mathbf{a}$ no longer changes between iterations within some small $\epsilon$:
\par
\item[1] Find $a_2$ such that $\alpha^\star_2=\alpha^\star_1$.
\par
$\vdots$
\item[H-1] Find $a_K$ such that $\alpha^\star_K=\alpha^\star_1$.
\par
\item[H] Find $a'$ such that $\mathbf{a}=a'(a_1,a_2,\hdots,a_K)$ results in Equation \eqref{FWERequation} with $\Theta=0$ equalling $\alpha$.
\end{itemize}
\label{Alg:FWER}
\end{algorithm}
\subsection{Power and sample size for each treatment}
The aim of the trial design in question is to find the required sample size to achieve the power for every hypothesis greater than a pre-specified level  $(1-\beta)$.  We assume that any given treatment $k'$,  is recommended when (i) its test statistic crossed the corresponding upper boundary, and (ii) its test statistic is the largest one, where $k' = 1,\hdots, K$. The power is defined as the probability of rejecting $H_{0k'}$ for $k'$, such that $\mu_{k'}-\mu_0=\theta'$ and $\mu_{k}-\mu_0=\theta_0$ for $k \neq k'$ where $\theta'$ is the clinically interesting treatment effect and $\theta_0$ is the highest uninteresting treatment effect. This setting is known as the least favourable configuration for treatment $k'$ \citep[which is denoted as LFC$_{k'}$,][]{THALLPETERF1988Tsat}. 
\par
Let $\Pi_{k',J'}$,  denote the probability that under the LFC$_{k'}$, no null hypotheses are rejected before the $J'$th analysis for treatment $k'$,  with treatment $k'$ not being stopped for futility at any of these analyses, and at analysis $J'$, $H_{0.k'}$ being rejected and treatment $k'$ being recommended, where $J' = 1,\hdots, J_{k'}$. The power for rejecting treatment $k'$ is then given by $\Pi_{k',1}+ \Pi_{k',2} \hdots + \Pi_{k',J_{k'}}$. To obtain $\Pi_{k',J'}$, we find the probability that $H_{0k'}$ is not rejected before analysis $J'$ and treatment $k'$ is not dropped for futility before analysis $J'$ assuming  $t_1,\hdots ,t_{s(k')+J'}$ and $v_{J'}$ are known, where $t$ is defined in Equation \eqref{equ:t} and $v_{J'}=\frac{\bar{X}_{k',J'}-\mu_{k'}}{ \frac{\sigma}  {\sqrt{n_{k',J'}}}}$, with $t_1,\hdots ,t_{s(k')+J'}$ and $v_{J'}$ and then integrate over every possible value as can be seen below.  This probability is
$
\Phi_{J'-1} (\tilde{L}_{k',J'-1}(\theta'), \tilde{U}_{k',J'-1}(\theta'), \tilde{\Sigma}_{k',J'-1}),
$
where $\tilde{L}_{k',J'-1}(\theta)=(\tilde{l}_{k',1}(\theta),\hdots \tilde{l}_{k',J'-1}(\theta))$ and $\tilde{U}_{k',J'-1}(\theta)=(\tilde{u}_{k',1}(\theta),\hdots \tilde{u}_{k',J'-1}(\theta))$. The $(i,i^\star)$th element $(i\leq i^\star)$ of the correlation matrix $\tilde{\Sigma}_{k',j}$ is $\sqrt{\frac{r_{k',i}(r_{k',J'}-r_{k',i^\star})}{r_{k',i^\star}(r_{k',J'}-r_{k',i})} }$ and
\begin{align*}
\tilde{l}_{k',j}(\theta)=& \sqrt{\frac{r_{k',J'}}{r_{k',J'}-r_{k',j}}} \bigg{(} l_{k',j}(\theta)-v_{J'} \sqrt{\frac{r_{k',j}}{r_{k',J'}}} \bigg{)},
\\
\tilde{u}_{k',j}(\theta)=& \sqrt{\frac{r_{k',J'}}{r_{k',J'}-r_{k',j}}} \bigg{(} u_{k',j}(\theta)-v_{J'}\sqrt{\frac{r_{k',j}}{r_{k',J'}}} \bigg{)}.
\end{align*}
The event that $H_{0k'}$ is  rejected at analysis $J'$ assuming that  $t_1,\hdots ,t_{s(k')+J'}$ and $v_{J'}$ are known and where $\mathbbm{1}\{ \cdot \}$ is an indicator function, is
\begin{align*}
\xi_{J'} = & \mathbbm{1} \{ \frac{v_{J'}}{\sqrt{n_{k',J'}}}- \frac{\sum_{i=1}^{J'} t_{s(k')+i} \sqrt{n_{0,s(k')+i}-n_{0,s(k')+i-1}}}{n_{0,s(k')+J}-n_{0,s(k')}}  +\frac{\theta'}{\sigma} > u_{J'} \sqrt{n^{-1}_{k',J'}+(n_{0,s(k')+J'}-n_{0,s(k')})^{-1}} \}, 
\end{align*}
\par
The probability that $H_{0k}$ is not rejected before analysis $J'$ for treatment $k'$ for all $k\in 1,\hdots,k-1,k+1, \hdots K$ assuming  $t_1,\hdots ,t_{s(k')+J'}$ and $v_{J'}$ are known is
\begin{align*}
\gamma_{k,J'}= &\mathbbm{1}\{ s(k')+J'-s(k)>0 \} \bigg{[} \sum_{j=1}^{\min( J_k,s(k')+J'-s(k) )}
 \mathbbm{1}\{s(k')+J' > s(k)+j\}  \Phi(L_{k,j}(\theta_0),U_{k,j}(\theta_0),\Sigma_{k,j}) 
\\
& + \mathbbm{1}\{s(f)+J' = s(k)+j\}  \Phi(L_{k,j}(\theta_0),\dot{U}_{k,j}(\theta_0),\Sigma_{k,j}) \bigg{]}+ \mathbbm{1}\{ s(k')+J'-s(k)\leq 0\},
\end{align*}
where $\dot{U}_{k,j}(\theta_0)=(u_{k,1}(\theta_0),\hdots u_{k,j-1}(\theta_0), \dot{u}_{k,j}(\theta_0)),$
with
\begin{align*}
\dot{u}_{k,j}(\theta_0)=& \max \bigg{[} u_{k,s(k')-s(k)+J'}(\theta_0), \frac{\sqrt{n_{k,s(k')-s(k)+J'}}}{\sqrt{n_{k',J'}}}v_{Y} +\frac{n_{k,s(k')-s(k)+J'}(\theta' -\theta_0)}{\sigma} \bigg{]}.
\end{align*}
One can then find $\Pi_{k',J'}$ as
\begin{align}
\begin{split}
\Pi_{k',J'}= & \int_{-\infty}^{\infty} \hdots \int_{-\infty}^{\infty}   \Phi(\tilde{L}_{k',J'-1}(\theta'),\tilde{U}_{k',J'-1}(\theta_1), \tilde{\Sigma}_{k',J'-1})
\\ & (\prod_{k=1 , k \neq k'}^K \gamma_{k,J'}) (\xi_{J'}) \; d\Phi(t_1) \hdots d\Phi(t_{s(k')+J'}) \; d\Phi(v_{J'}).
\end{split}
\label{eq:power}
\end{align}
\par
To ensure that all the  experimental treatments achieve the pre-specified power under the corresponding LFC$_{k'}$, the sample size must be found in order for $\Pi_{k',1}+ \Pi_{k',2} \hdots + \Pi_{k',J_{k'}}  \geq 1-\beta$ for all $k^\prime$. Due to treatments potentially starting at different times and having different number of stages, each treatment may require a different number of patients to achieve the same power. As changing the sample size of one treatment effects the power of another, an iterative approach is proposed to calculate the required sample size per treatment per stage. Specifically, to have all the treatments controlled at the same specified power, $1-\beta$ under their LFC$_{k'}$, one needs to define $\mathbf{n}= (n_0,\hdots, n_K )$, then by assuming each $n_k$ can take any real value,  use Algorithm \ref{Alg:Power}. Once $\mathbf{n}$ is found using this Algorithm \ref{Alg:Power}, round up each value of $\mathbf{n}$ to its nearest integer, then recalculate $\mathbf{a}$ with these new values for $\mathbf{n}$ to account for the fact that the ratios have now also changed between treatments. 
\begin{algorithm}[]
\caption{Iterative approach to compute the sample size}
\begin{itemize}
\item[0] Begin by assuming $n_k=1$ for all $k, \in 0,\hdots,K$ then calculate $\mathbf{a}$ using Algorithm \ref{Alg:FWER}. Find $n_1$ such that $\Pi_{1,1}+ \Pi_{1,2} \hdots + \Pi_{1,J_1}=1-\beta$ with $n_k=n_1$ for all $k \in 0,\hdots,K$  and update $\mathbf{n}$. Then repeat the following iterative steps until each element of $\mathbf{n}$ no longer changes between iterations within some small $\epsilon$:
\item[1] Find $n_1$ such that $\Pi_{1,1}+ \Pi_{1,2} \hdots + \Pi_{1,J_1}=1-\beta$. 
\item[2] Find $n_2$ such that $\Pi_{2,1}+ \Pi_{2,2} \hdots + \Pi_{2,J_2}=1-\beta$.
\par
$\vdots$
\item[H] Find $n_K$ such that $\Pi_{K,1}+ \Pi_{K,2} \hdots + \Pi_{K,J_K}=1-\beta$.
\item[H+1] Find $n_0$ and $r_{0,1},\hdots r_{0,J_0}$ based on $n_1,\hdots,n_K$ then recalculate $\mathbf{a}$ using Algorithm \ref{Alg:FWER}.
\end{itemize}
\label{Alg:Power}
\end{algorithm}
\subsection{Sample size distribution and Expected sample size}
\label{sec:expsam}
The distribution of the sample size and expected sample size can both be calculated by finding the probability of every possible outcome of the trial denoted by $P_{\tilde{J},Q}$. Define $\tilde{J}=( \tilde{j}(1),\hdots, \tilde{j}(K))$ with $\tilde{j}(k) = {1,\hdots, J_k}$ as the point in which treatment $k$ would finish being tested, ignoring the possibility that the trial has already stopped early as a different treatment is found which is superior to the control. This is done in order to remove the dependence between each active arm. We define $Q=( q(1),\hdots, q(K) )$ with $q(k)= \infty$ if treatment $k$ goes below the lower stopping boundary at point $\tilde{j}(k)$ and $q(k)= 1$, if treatment $k$ goes above the upper stopping boundary at point $\tilde{j}(k)$. Due to ignoring the possibility of the trial already stopping early, every active treatment will either stop for futility or efficacy therefore $q(k)$ can only take one of two values. We find
\begin{align*}
P_{\tilde{J},Q}= & \int_{-\infty}^{\infty} \hdots \int_{-\infty}^{\infty} \prod_{k=1}^K \Bigg{[} \mathbbm{1}\{q(k)=1\} \Phi(L^{+}_{k,\tilde{j}(k)}(\theta_k),U^{+}_{k,\tilde{j}(k)}(\theta_k), \Sigma_{k,\tilde{j}(k)}) 
\\ & + \mathbbm{1}\{q(k)= \infty\} \Phi(L_{k,\tilde{j}(k)}(\theta_k),U_{k,\tilde{j}(k)}(\theta_k),\Sigma_{k,\tilde{j}(k)})  \Bigg{]}  
 d\Phi(t_1), \hdots, d\Phi(t_{\max(\tilde{J}+S)}),
\end{align*}
where $U^{+}_{k,j}= (u_{k,1}(\theta_k),\hdots u_{k,j-1}(\theta_k), \infty)$ and $ L^{+}_{k,j}= (l_{k,1}(\theta_k),\hdots l_{k,j-1}(\theta_k), u_{k,j}(\theta_k))$. The $P_{\tilde{J},Q}$  are then associated with their given total sample size $N_{\tilde{J},Q}$ for that given $\tilde{J}$ and $Q$.
\begin{align*}
N_{\tilde{J},Q} = &\bigg{(} \sum^K_{k=1} n_{k,\max(\min(\tilde{j}(k)+s(k),(\tilde{J}+S)\circ Q)-s(k),0)} \bigg{)}
+n_{0,\min(\max(\tilde{J}+S),(\tilde{J}+S)\circ Q)},
\end{align*}
where $\circ$ is the scalar product. To obtain the sample size distribution each value of $\tilde{J}$ and $Q$ which result in the same value of $N_{\tilde{J},Q}$ is associated with its corresponding $P_{\tilde{J},Q}$. This set of $P_{\tilde{J},Q}$ is then summed together to give the probability of the realisation of this sample size. 
To find the sample size distribution for each active arm one can associate $n_{k,\max(\min(\tilde{j}(k)+s(k),(\tilde{J}+S)\circ Q)-s(k),0)}$ with its corresponding $P_{\tilde{J},Q}$, and this can similarly be done for the control treatment. 
The expected sample size for N for a given $\Theta$, $E(N|\Theta)$, can be found by summing up every possible combination of $\tilde{J}$ and $Q$,
\begin{equation*}
E(N|\Theta)= \sum_{\substack{\tilde{j}(k)=1 \\ k=1,2,\ldots,K}}^{J_k} \sum_{\substack{q(k)\in \{ 1,\infty \} \\ k=1,2,\ldots,K}} P_{\tilde{J},Q} N_{\tilde{J},Q}.
\end{equation*}

\section{Numerical Evaluations}
\label{sec:Example}
\subsection{Motivating trial}
In recent years, there have been several platform trials conducted and their use appears to be increasing during the COVID-19 pandemic \citep{StallardNigel2020EADf}.
One platform trial example is FLAIR \citep{HowardDenaR2021Apti}. It is a randomised, controlled, open-label, confirmatory trial in chronic lymphocyte leukaemia.
When designing  FLAIR  there was the plan to add an active treatment during the trial as well as an interim analysis halfway through the planned sample size for each treatment. In the actual trial, two additional arms were added, one being an additional control arm. The original design of the study considered pairwise type I error control only and hence did not provide the FWER control. 
\par
We revisit the design of the FLAIR trial accounting for one additional active treatment arm being planned to be added mid-trial. One active and a control arm begin the trial, and apply the proposed methodology to control the FWER in the strong sense. One can argue that the controlling FWER in this trial is essential as the trial aimed to test different combinations of treatments with the same common compound - ibrutinib for all the active treatments whilst the main control was not based on ibrutinib \citep{WasonJamesMS2014Cfmi}. This might be a regulatory requirement to control the FWER while adding an arm during the trial \citep{FDA}.
\par
Based on the planned effect given in FLAIR, the interesting treatment difference is $\theta'=-\log(0.69)$, $\sigma=1$, and the uninteresting treatment effect is assumed (as not being used in the original design) to be $\theta_0=-\log(0.99)$. 
The desired power in FLAIR was 80\%, while the type-I error of each treatment comparison was 2.5\% (one-sided). While still targeting the same power, we will use a more stringent target of 2.5\% FWER (one-sided). In line with FLAIR, we use a total number of stages for both settings to be three. Therefore, in Setting 1, treatments 1 and 2 will both have two stages, whereas, in Setting 2, Treatment 1 will have three stages and Treatment 2 will have two (see Figure \ref{fig:screp}).
The interims are equally spaced for the active treatments across all stages so $r_{k,j}=j$ for $k > 0$. Informed by the recruitment to FLAIR, we assume a constant recruitment rate of 21 patients per month. 
\par
The operating characteristics that will be studied for both methods include the FWER and power under LFC$_k$. These two are studied to ensure that the trial design meets the required error control. Other operating characteristics stated include the maximum number of stages per active treatment arm, denoted by NS$_k$, as well as the number of patients per arm per stage. Also shown for each setting is the maximum sample size and duration until the trial is complete as well as the expected sample size and duration. The duration of the trial is denoted by $T$. These values are found in order to compare the different designs.
\subsection{Competing designs}
We compare the proposed designs to three alternative methods. All of these are in the frequentest framework. For brevity, the main focus of the competing designs comparison will be on the Setting 2 (with the results for Setting 1 provided in Supporting Information).

The first competing approach is to evaluate each active treatment in separate trials. In line with Setting 2, the first study uses the 3-stage design and the second one - the 2-stage one.  This approach will be referred to as ``Separate trials''. Two variations on running separate trials are studied. The first controls the FWER across both trials using $\alpha'$, the error rate for each trial, chosen so that $(1-\alpha')^2=0.975$. This results in a type I error for each trial of 1.26\%. The second variation does not control the FWER across the two trials.
\par
The second competing design is the MAMS approach proposed by~\cite{MagirrD.2012AgDt}. We will refer to these approach as ``MAMS trial''. Note, that under this MAMS approach, the trial cannot start until all treatments are ready.
As this approach requires equal numbers of stages per treatment, both the results for running a 2 stage and 3 stage trial are presented. 
\par
The third competing method uses the same design parameters as originally planned for the 2-arm 3-stage trial and then also uses them for the additional treatment. This approach will be referred to as the ``Naive MAMS'' as it does not adjust the design parameters for the added arm. This will also allow to demonstrate the effect of not adjusting a design for additional arms. We provide the results for both the same maximum sample size as originally planned, and for the same sample size per arm per stage, $n_{j,k}$.
\par
The operating characteristics of the different design options are provided in Table \ref{tab:setting12}. The calculations were carried out using R \citep{Rref} with the method given here having the multivariate normal probabilities being calculated using the package \texttt{mvtnorm} \citep{mvtnorm} and the outer integrals being calculated using the quadrature rule with the packages \texttt{gtools} \citep{warnes2021package} and \texttt{statmod} \citep{smyth2021package}. Code is available at \textit{https://github.com/pgreenstreet/platform-design-with-addition-of-arms}. The comparison results were obtained using the R package \texttt{MAMS} \citep{jaki2019r}. 

\subsection{Results}
\label{sec:mainresults}
All the results can be seen in Table \ref{tab:setting12}. For all the designs the triangular shaped stopping boundaries are used \citep{WhiteheadJ.1997TDaA,WasonJamesM.S.2012Odom}. The first two rows of Table \ref{tab:setting12} shows the proposed design under Setting 1 and 2, respectively with the Supporting Information containing the results of using other stopping boundary shapes.

\begin{table}[H]
\centering
\caption{Operating characteristics of the proposed design under two settings and competing approaches: Running trials separate (``Separate Trials''), MAMS design  by \citep{MagirrD.2012AgDt} (``MAMS''), and using a ``naive'' MAMS approach for the FLAIR trial.}
\begin{tabular}{c|c|c|c|c|c|c|c|c}

& FWER & LFC$_1$ &   NS$_1$  & $n_1$ & $\max(N)$ & $E(N|H_{G})$ & $E(N|\text{LFC}_1)$ & $E(N|\text{LFC}_2)$ \\
&  & LFC$_2$ &  NS$_2$  &  $n_2$ & $\max(T)$ & $E(T|H_{G})$ & $E(T|\text{LFC}_1)$ & $E(T|\text{LFC}_2)$ \\
\hline

Setting 1 & 0.025 & 0.802  & 2 & 76   & 540  & 351.8  & 285.8  &  400.8  \\
&  & 0.804 & 2   & 78  &  (25.7) &  (16.8) &  (13.6) &   (19.1) \\
\hline
Setting 2 & 0.025 & 0.802  & 3  & 46  & 492  & 303.3  & 296.6 &  347.8  \\
 &  &  0.803 &  2   & 77  & (23.4) &  (14.4) & (14.1)&   (16.8)  \\
\hline
Separate trials   & 0.025 & 0.801 & 3   &  53  & 626  & 349.1  &  405.0  & 400.6   \\
FWER control & & 0.805  & 2  &  77 &  (29.8) &  (16.6) &  (19.3) &  (19.1)  \\

\hline
Separate trials & 0.049 & 0.807 & 3   & 46  & 536  & 302.2  & 340.1  & 336.1  \\

no FWER control & & 0.803 & 2  & 65 &  (25.5) &  (14.4) &  (16.2) &  (16.0) \\

\hline
MAMS trial &0.025 & 0.804 & 2   & 76  & 456  & 280.7  & 309.8  & 309.8\\
2 Stage & & 0.804 & 2  & 76 & (26.1) &  (17.7) &  (19.1) &  (19.1) \\
\hline
MAMS trial  & 0.025 & 0.805 &3   &  53  & 477  & 258.0 & 289.4  & 289.4  \\
3 stage & & 0.805 &3  &  53 & (27.1) &  (16.7) &  (18.2) &  (18.2) \\
\hline
Naive MAMS  & 0.044 & 0.804  & 3   & 46  & 368  & 219.0  & 217.3  & 239.7 \\
same $n_{j,k}$ &  &  0.564  &2   & 46 &  (17.5) & (10.4) & (10.3) &  (11.4) \\
\hline
Naive MAMS  & 0.044 & 0.716  & 3   &  46/30  & 276  & 177.2  & 178.2  & 190.8 \\
same $\max(N)$ & &  0.408 & 2   & 31 &  (13.1) &  (8.4) & (8.5) &  (9.1) \\
\hline
\end{tabular}
\\
\vspace{0.1cm}
Key: $E(N|H_{0.G})$, $E(N|\text{LFC}_k)$, $E(T|H_{0.G})$, $E(T|\text{LFC}_k)$ is the expected sample size and trial duration under the null and under the LFC for treatment $k$, respectively. 
\label{tab:setting12}

\end{table} 

Under Setting 1, which has two stages for each active treatment, the design requires 42 and 44 patients per stage for Treatment 1 and Treatment 2, respectively. As a result, the maximum sample size is 540, and the expected sample sizes varies between approximately 286 and 401. The maximum duration of the trial is 25.7 months, and the expected one varies between 13.6 and 19.1 months. Under Setting 2, in which the first active treatment has 3 stages and the second has only 2, the required sample size per stage are is 46 and 77, respectively. This results in the maximum sample size of 492 and expected sample size between 296.6 and 347.8 depending on the configuration. This resulted in a maximum duration of 23.4 months and the expected duration varies between 14.1-16.8 months.
\par
Comparing the sample size and trial duration for the proposed designs under Settings 1 and 2, under most configurations Setting 2 has lower expected sample sizes, and requires nearly 50 fewer patients in the maximum sample size to achieve 80\% power while controlling the FWER at 2.5\%. This also translates into the shorter duration. Setting 1 has advantages over Setting 2 under the case when Treatment 1 is superior, and Treatment 2 has the uninteresting effect. However, the difference in the expected sample size is around 10 patients and the difference in the expected duration is 0.5 months. For this reason, we will focus on the comparisons with Setting 2 with the results for Setting 1 provided in Supporting Information.

\par
The next two rows of Table \ref{tab:setting12} show the operating characteristics of running two separate trials. To match the setting of the FLAIR (and Setting 2), it is assumed that the first treatment has three stages while the second has two. Under the FWER across these separate trials controlled, the maximum and expected sample sizes are noticeably larger - with the difference of 134 patients required to achieve 80\% power. Running two separate trials with the FWER also increases the maximum duration by 6 months, and the expected duration by 2-5 months, on average, depending on the configuration. Under two trials not controlling the FWER, the advantage of the proposed design under Setting 2 still persists. Two separate trials would require 44 more patients, and the expected sample size are only nearly 11 patients lower under LFC$_2$ which results in less than 1 month in recruitment time. Comparing this to the expected sample size under LFC$_1$ which is around 44 patients lower for Setting 2 which results in a saving of time of over 2 months. 
\par
The second competing method is the MAMS approach proposed by \cite{MagirrD.2012AgDt} that requires all treatments to start at the same time, and uses the critical values controlling the FWER, and the sample size achieving 80\% power. We consider both 2- and 3-stage variants for a fairer comparison. The duration until the trial finishes for both includes the time before the trial would be able to start. This is calculated by assuming that Treatment 2 is ready when planned for Setting 2 and then calculating the time before this treatment is added. Therefore, this is the time for the first 92 patients to be recruited - 4.4 months.  
Under this MAMS design, the maximum sample size is lower than for the proposed one under Setting 2 (36 and 15 patients for the 2- and 3-stage designs, respectively). The maximum duration, however, is increased by 3-4 months. The expected duration is also  increase for all the configuration studied with an increase of between 1.4-5 months. 
\par
The final comparison method of naively using the original design for a 2 arms 3 stage trial is shown. The original design is to have 46 patients per arm per stage which results in 276 patients. Therefore when $n_{j,k}$ is kept the same this results in a maximum sample size of 368. For this approach the FWER is inflated by over 75\%. The power under LFC$_1$ is still above the desired level however under  LFC$_2$ the power decreases to 56.4\% so is well below the target of 80\%.    
For the naive approach where the maximum sample size remains the same there is therefore a change in sample size for the first treatment's 2$^\text{nd}$ and 3$^\text{rd}$ stages to accommodate the addition of the new treatment. As a result there is 46 patients on Treatment 1 at stage 1 then this decreases to 30 for Treatment 1's final two stages. In order to keep $\max(N)=276$ then $n_2$ was set to equal 31 patients. In this case the FWER is inflated by over 75\% and neither the power under the LFC$_1$ or LFC$_2$ is controlled at the desired level. The drop in power for the LFC$_2$ between Setting 2 and this naive approach is over 35\% which is a dramatic loss in power. This poor result is to be predicted for this naive approach as it does not have bounds designed to control FWER or the required number of patients to get the desired power for either treatment.  
\par

\par
The results of other common stopping boundary shapes and combinations of these can be seen in the Supporting Information in Web Appendix D for Setting 1 and Setting 2. In Web Appendix C of the Supporting Information the comparison results to Setting 1 can be seen when using the triangular stopping boundaries.

Overall this section has shown how the methods proposed in this paper could work in order to design a clinical trial in which an additional treatment is added later. In addition, competing frequentest  approaches are studied to see how they compare. It can be seen that there is benefit to using the method proposed here compared to using these other methods either with regards to sample size or trial duration. 

\subsection{Sample Size Distribution}
\begin{figure}[H]
\begin{subfigure}{1\textwidth}
  \centering
  \includegraphics[width=.9\linewidth,trim= 0 0.5cm 0 1cm, clip]{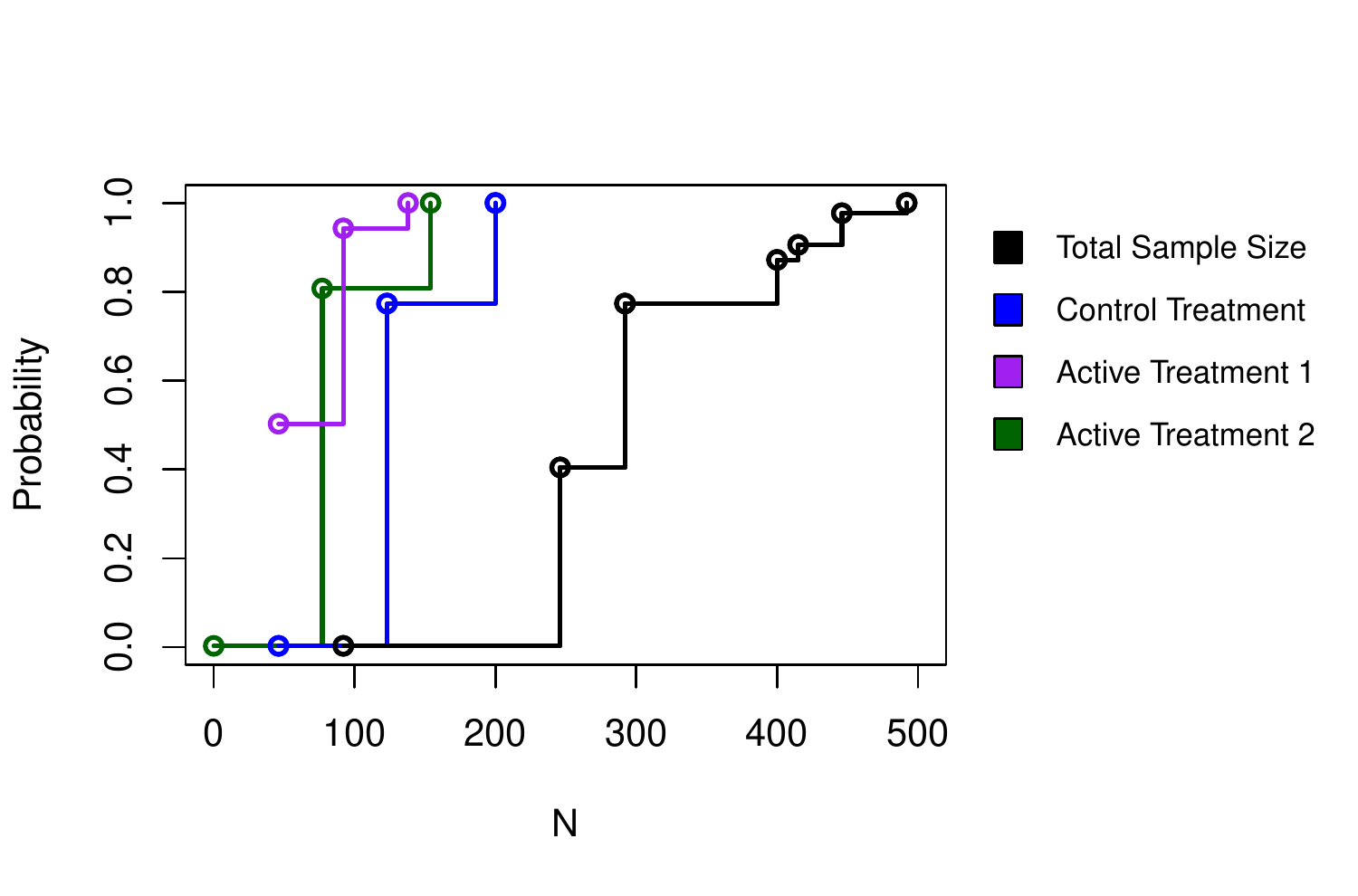}  
\end{subfigure}
\caption{Cumulative distribution functions (CDF) of the number of treatments needed for the trial in Setting 2 of the total sample size and for each arm individually}
\label{fig:3distribution}
\end{figure}

The distribution of the total sample size and the sample size of each treatment for Setting 2 under the global null is given in Figure \ref{fig:3distribution}. Analogous results for Setting 1 can be seen in the Supporting Information (Web Appendix E) along with the expression for the probability mass function for the total sample size for Setting 2. The design under Setting 2 results in the interquartile range of 246 to 292 and median of 292 under the global null for the total sample size. These figures can be used by the trial team and given to funders and regulators to help with the communication of how many patients are likely to be required for the trial.   
\par
\section{Robustness to deviations in the planned adding}
\label{sec:sim}
In the FLAIR trial, the second active treatment was not added until about three quarters of the way through the recruitment for the first treatment. In this section, the effect of adding the treatments earlier or later than planned (i.e. at the first interim for the considered example) will be studied using simulations. We consider three approaches of how a treatment could be added later or earlier to a trial are studied. In all approaches the total maximum sample size is fixed to be the same. Below, we focus on Setting 2, and similar results for Setting 1 are given in the Supporting Information Web Appendix F.
\par
Approach 1 is to change the timing of the interim analysis for Treatment 1 so it is conducted when Treatment 2 is added. Once the second active treatment is added, the allocation ratio for Treatment 1 to control changes as  in the original proposed design. The patients remaining from the total sample size are then shared out across the phases with respect to the pre-set allocation ratio between each treatment. The pre-set stopping boundaries are used. Approach 2 follows Approach 1, but instead of keeping the original boundaries the bounds are recalculated using Algorithm \ref{Alg:FWER} with the allocation ratios of each treatment at the time the additional treatment is actually added. Approach 3 keeps the timing of the interim analysis for Treatment 1 unchanged, and, at this point, the allocation ratio changes.
 
The effect of adding the second active treatment to the trial after only recruiting 1 patient to the control, up to recruiting 189 patients to the control, is studied. With the first treatment receiving the correct number of patients based on its recruitment rate relative to the controls recruitment rate. i.e 1 patient will have also been recruited to Treatment 1 before Treatment 2 is added in the earliest example. 
\par
Figure 3 shows the resulting FWER and Power for different times when the new treatment is added based on 10 million simulations for each case. 
\begin{figure}[]
\begin{subfigure}{1\textwidth}
  \centering
  \includegraphics[width=.7\linewidth,trim= 0 0cm 0 1cm, clip]{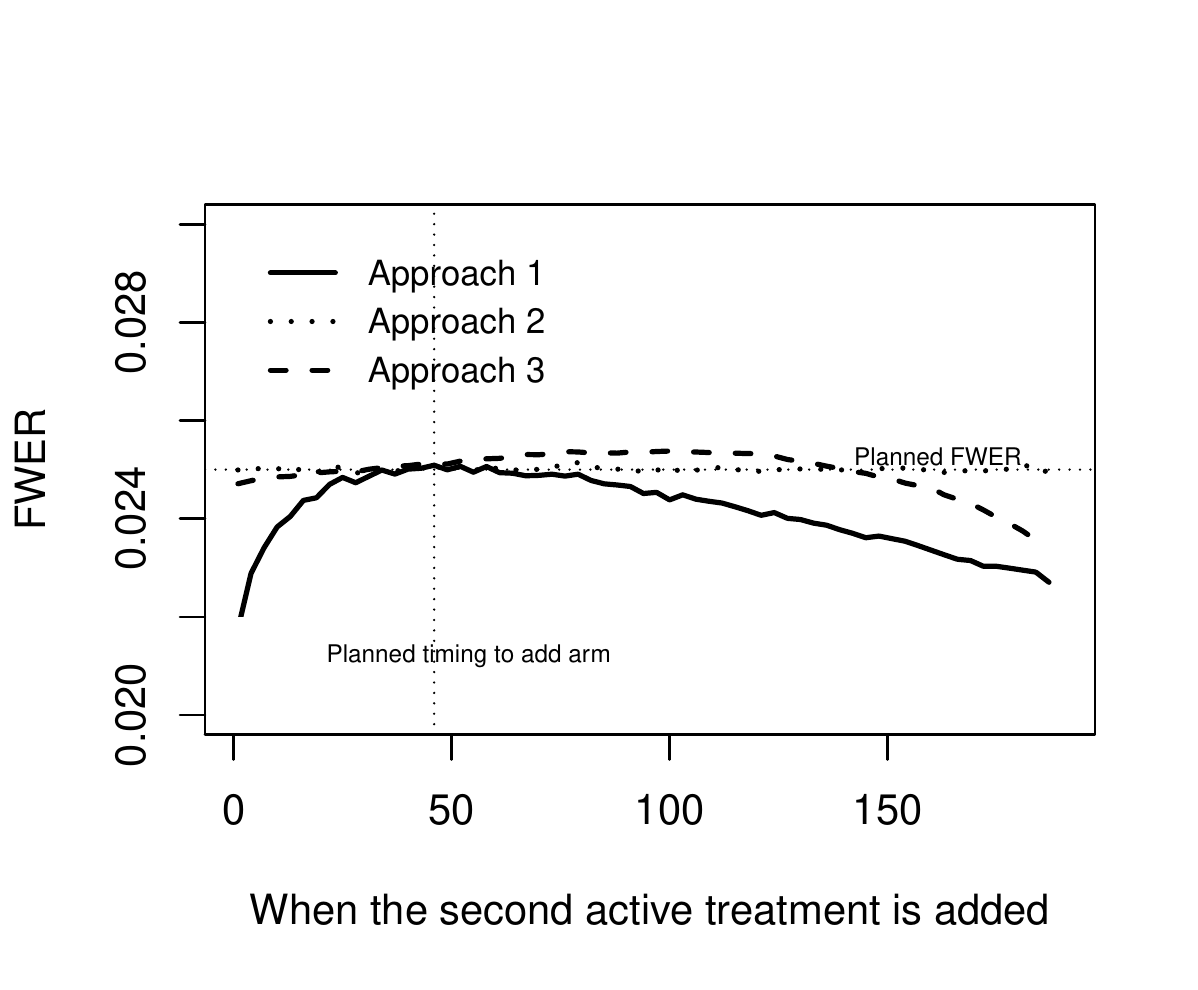}  
  \caption{}
  \label{fig:3stagesima}
\end{subfigure}
\begin{subfigure}{1\textwidth}
  \centering
  \includegraphics[width=.7\linewidth,trim= 0 0cm 0 1cm, clip]{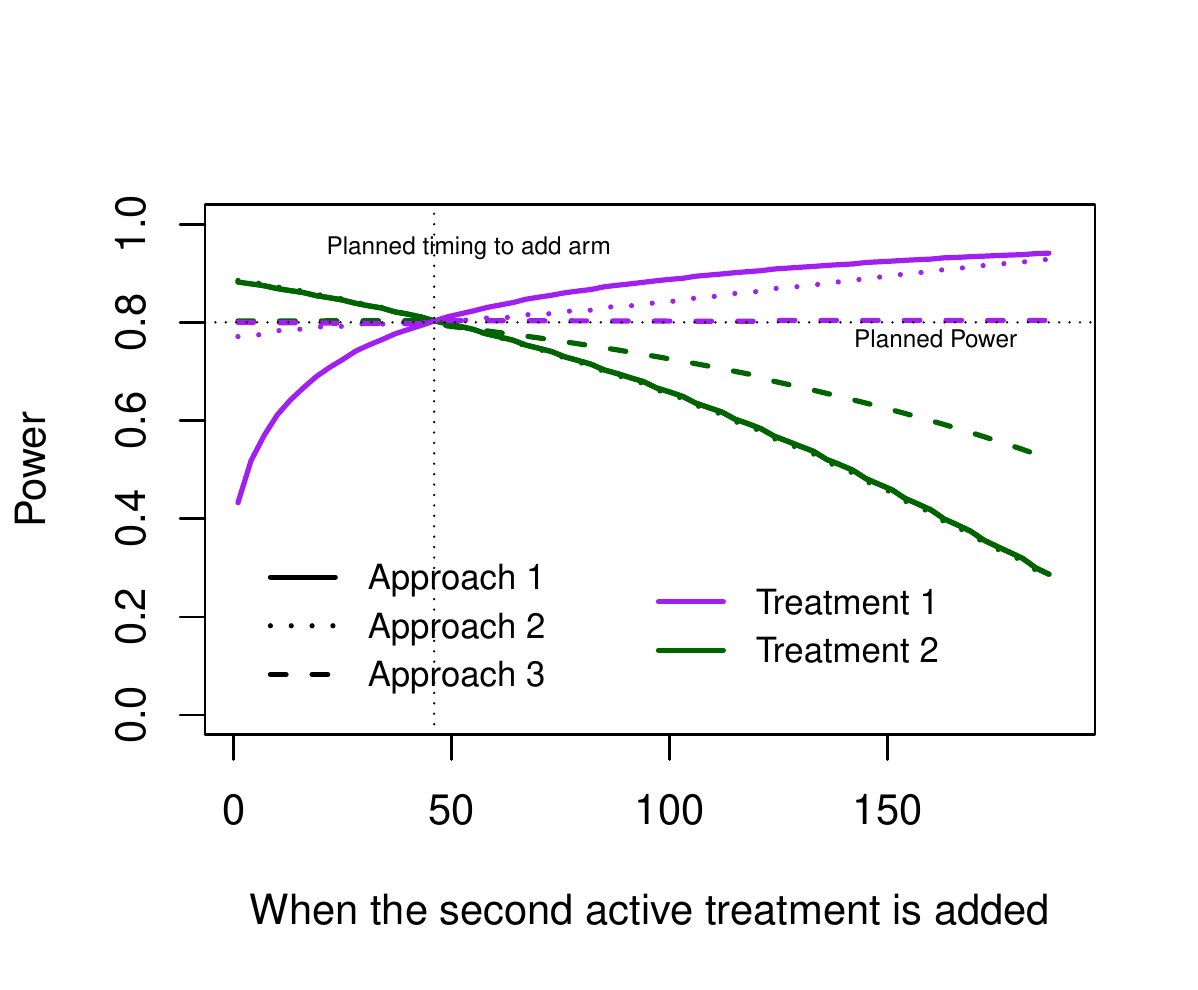}  
  \caption{}
  \label{fig:3stagesimb}
\end{subfigure}
\caption{The effect of adding the treatment later or earlier than planned using three different approaches for Setting 2 on power and FWER. With sub-figure (a) showing the FWER under the global null, and in sub-figure (b) showing the power under the LFC for Treatment 1 and the power under the LFC for Treatment 2.}
\label{fig:3stagesim}
\end{figure}

Figure \ref{fig:3stagesima} shows how the FWER varies for each one of the approaches with Approach 2 having the least variation and Approach 1 having the most variation in FWER under the global null. The maximum inflation happens for Approach 3 of an increase in FWER under the global null to 2.54\% when the second treatment is added after 100 patients are recruited to the control. For Approach 1 the FWER increases until the planned adding time and then decreases. Approach 2 stays constantly around the planned FWER and 
 for Approach 3 the FWER starts below 2.5\% and increases until 100 patients, before starting to decrease.
\par
The changing FWER for Approach 1 and 3 are caused by two opposing forces. The first of which is when Treatment 1 is added earlier  there is a decrease in correlation between the first interim for Treatment 1 and the rest of its analyses, this is caused by a decrease in $\sqrt{r_{1,1}/r_{1,2}}$ and $\sqrt{r_{1,1}/r_{1,3}}$. The second, is there is an increase in correlation between the Z-statistics for Treatment 1 and 2 as there is now an increased number of shared control patients. These two opposing forces make it difficult to predict what effect any change will have on the FWER without running the calculations or using simulations. In order to guarantee that the FWER is controlled then either the second treatment needs to be added when it was planned to be or recalculate the stopping boundaries for each point as done in Approach 2. 
\par

Considering the power, for the all considered approaches the later the additional arm is added the lower the power for this arm is. For Approaches 1 and 2, the power for the Treatment 1 increases the later the additional arm  is added whereas the power remains almost constant for Approach 3. One issue therefore with Approaches 1 and 2 is that power for all the treatments under the LFC is no longer controlled unless the treatment is added at the preplanned time. However, when using Approach 3, the power is controlled for both treatments when the treatment is added earlier as well as the FWER being controlled. This is not always the case as can be seen in the Supporting Information (Web Appendix G) which provides an example with higher uninteresting treatment effect. Higher $\theta_0$ results in a greatly increased chance of taking Treatment 2 forward before the second analysis for Treatment 1 due to $\theta_0$ effect on the sample size of the second active treatment. This reiterates why using the pre-planned design and assessing the impact of deviations of the plan are crucial. One potential solution to this problem of controlling the 
errors when adding the treatment at a different time to when it was planned is to recalculate all design parameters (including the sample size).
\par

This section has highlighted the importance of using the original plan in order to control the FWER and power under the LFC. Therefore it is important when using this design to ensure that the additional treatment will be ready for the pre-planned addition  time. One key point is that if the new treatment is not added to the trial at all, the design will still guarantee control of FWER and power for the treatments already in the trial. Furthermore by using the original plan this removes the bias potentially caused by changing when the additional treatments are added, in order to benefit treatments already in the trial. 


\section{Discussion}
\label{sec:discussion}
In this paper, a general design for adding additional treatments in a  pre-planned manner is developed and explored. This design ensures strong control of the FWER and power under the least favourable configuration and allows for interim analyses. Both sample size
distribution and expected sample size can be calculated. Two iterative approaches are given to allow for multiple stopping boundary shapes and to allow for different numbers of patients on each treatment depending on when the treatment is added to the trial. Two different designs based on FLAIR for the two settings are presented. These designs are then further explored where the effect of adding treatments later or earlier than planned is studied.
\par
Overall the method proposed here, which builds on the work of \cite{MagirrD.2012AgDt}, has shown that running a platform trial where treatments are added at later points can result in a considerably more favourable design to running separate trials with respect to maximum and expected sample size. This approach has shown that it can be worthwhile starting a trial earlier with the available treatments and then planning to add treatments later, compared to waiting until all are ready then beginning the trial with respect to the time it takes before the trial concludes. This is true both when there is either a constant or increasing recruitment rate.  In Section \ref{sec:mainresults} it was seen that using Setting 2 compared to 1 can be potentially beneficial with regards to the trials sample size and duration. This makes intuitive sense as this results in increased correlation between the test statistics as there are more shared controls which results in a reduction in the FWER of the trial for the given boundaries. 
\par
Throughout this paper only concurrent controls are used, as argued by \cite{lee2020including}. 
While it would be possible to extend the design to allow for non-concurrent controls, error control will only be guaranteed if there is no difference between the concurrent and the non-concurrent controls, but this is impossible to know before the trial. In this paper we assumed  that an interim analysis is conducted at the time that a new treatment is added. This not only simplifies calculations, but is also sensible as if a new treatment is being added to the trial then the other treatments in the trial may as well also be studied at this point. This has two benefits, the first is there is the opportunity for a treatment to be declared superior to the control before recruitment of the new treatment starts. The second is it allows the study of all the patients on the control treatment from before the additional treatment is added, so potentially making it easier at later stages to know which controls are concurrent.  
\par
PWER was used in order to calculate  $a_1,a_2,\hdots,a_K$ in order to share the FWER out among the treatments.  This was used as it ensures the highest possible probability of rejecting a null hypothesis is the same for all treatments. However there are a multitude of different ways the FWER could be shared such as having: the probability of rejecting a null hypothesis under the global null the same; or the probability that a treatment is taken forward to the next phase the same. One may also want to consider one of these approaches. To do this the same iterative approach as given in Section \ref{FWERsection} with a different Equation \eqref{PWER} can be used.
\par
When calculating the expected sample size every possible outcome of the trial was enumerated resulting in a very computationally costly procedure. In Section Web Appendix B of the Supporting Information a more efficient approach is provided for the computation of the expected sample size.  The cost of this efficiency is that the algorithm does not yield the full sample size distribution in addition to the expected sample size.
\par  
An area for further research is deciding whether to wait for all the treatments to be ready or to start the trial with the ability to add preplanned treatments later. A lot of factors need to be considered when choosing this such as: recruitment time, recruitment cost, time left before all the treatments are ready, and the cost of delaying development of existing treatments. Therefore an area for further research from this paper is looking at how a decision framework, such as the one discussed in \cite{LeeKimMay2019Taon}, could be used. 
\par
Some of the potential limitations with this approach are that it assumes normally distributed data. By using asymptotic normality as discussed in \cite{JakiT2013Coca}, however, other endpoints can also be used. Another area is the fact it is assumed that the common variance is known. However using an ad hoc approach such as the one in \cite{MagirrD.2012AgDt} can also be used to transform individual test statistics to combat this issue.

\section*{Acknowledgements}

This report is independent research supported by the National Institute for Health Research (NIHR300576). The views expressed in this publication are those of the authors and not necessarily those of the NHS, the National Institute for Health Research or the Department of Health and Social Care (DHSC). PM and TJ also received funding from UK Medical Research Council (MC\_UU\_00002/14). 
This paper is based on work completed while PG was part of the EPSRC funded STOR-i centre for doctoral training (EP/S022252/1).\vspace*{-14pt} 

\newpage
\bibliography{refp1} 
\newpage
\beginsupplement
\section*{Supporting Information}
\section{Proof of strong control of FWER}
\label{App:Proof}
Recall that $A_{k,j}(\theta_k)=[Z_{k,j}<  l_{k,j} + (\mu_k-\mu_0 -\theta_k)I^{1/2}_{k,j}]$ and $B_{k,j}(\theta_k)=[ l_{k,j}  + (\mu_k-\mu_0 -\theta_k)I^{1/2}_{k,j} < Z_{k,j} <  u_{k,j} + (\mu_k-\mu_0 -\theta_k)I^{1/2}_{k,j}]$.

\begin{proof}
For any $\epsilon_k>0$,
\begin{align*}
\bigcup^{J_k}_{j=1}  \Bigg{[} \bigg{(} \bigcap^{j-1}_{i=1}B_{k,i}(\theta_k+\epsilon_k) \bigg{)} \cap A_{k,j}(\theta_k +\epsilon_k) \Bigg{]} \subseteq 
\bigcup^{J_k}_{j=1} \Bigg{[} \bigg{(} \bigcap^{j-1}_{i=1}B_{k,i}(\theta_k) \bigg{)} \cap A_{k,j}(\theta_k) \Bigg{]}.
\end{align*}
Take any
$$w=(Z_{k,1},\hdots,Z_{k,J}) \in \bigcup^{J_k}_{j=1}  \Bigg{[} \bigg{(} \bigcap^{j-1}_{i=1}B_{k,i}(\theta_k+\epsilon_k) \bigg{)} \cap A_{k,j}(\theta_k +\epsilon_k) \Bigg{]}.$$
For some $q \in \{1,\hdots, J_k \},$ for which $Z_{k,q} \in A_{k,q} (\theta_k +\epsilon_k)$ and $Z_{k,j} \in B_{k,j} (\theta_k +\epsilon_k)$ for $j=1,\hdots,q-1$. $Z_{k,q} \in A_{k,q} (\theta_k +\epsilon_k)$ implies that $Z_{k,q} \in A_{k,q} (\theta_k)$.
Furthermore $Z_{k,q} \in B_{k,q} (\theta_k +\epsilon_k)$ implies that $Z_{k,q} \in B_{k,q} (\theta_k) \cup A_{k,q} (\theta_k)$ for some $j=1,\hdots, q-1$.
Therefore,
\begin{equation*}
w \in \bigcup^{J_k}_{j=1} \Bigg{[} \bigg{(} \bigcap^{j-1}_{i=1}B_{k,i}(\theta_k) \bigg{)} \cap A_{k,j}(\theta_k) \Bigg{]}.
\end{equation*}

Next suppose for any $m_1, \hdots, m_K$ where $m_1 \in \{1,\hdots, K\} $ and $m_k \in \{1,\hdots, K\} \backslash \{ m_1,\hdots, m_{k-1} \}$ with $\theta_{m_1}, \hdots, \theta_{m_l} \leq 0$ and $\theta_{m_{l+1}}, \hdots, \theta_{m_K} > 0$. Let $\Theta_l=(\theta_{m_1},\hdots, \theta_{m_l})$. Then
\begin{align*}
P(& \text{reject at least one true } H_{0.k} | \Theta)
\\
& \leq P(Z_{k,j} > u_{k,j}  \text{ for some } (k,j)  \in \{ (m_1,1) \hdots, (m_1,J_{m_1}),(m_2,1)
\\
& \hdots,(m_l,1),\hdots, (m_l,J_{m_l}) \} | \Theta )
\\
& = 1- P(\bar{R}_l (\Theta_l))
\\
& \leq 1- P(\bar{R}_l (0)) 
\\
& \leq 1- P(\bar{R}_K (0)) 
\\
&= P(Z_{k,j} >  u_{k,j}  \text{ for some } (k,j) \in  \{ (m_1,1) \hdots, (m_1,J_{m_1}),(m_2,1) 
\\
& \hdots,(m_K,1), \hdots, (m_K,J_{m_K}) \} | H_{0.G})
\\
&= P(\text{reject at least one true } H_{0.k} | H_{0.G}).
\end{align*}
\end{proof}
  
\section{ Efficient computation of expected sample size}
The sample size calculation can be split into four sections. Two sections that focus on the control treatment and two sections that focus on the active treatments. These sections are:
\begin{enumerate}
\item The probability the control treatment finishes at each stage $j_0$ as the trial is stopped, given no null hypotheses are rejected, where $j_0 \in 1,\hdots, J_0$.
This is calculated by taking the difference between the probability that every treatment is stopped for futility by the control's $j_0$th stage, denoted by $\Psi_j$ and every treatment is stopped for futility by the control's stage $j_0-1$. The control treatment cannot stop being recruited until either: one or more null hypotheses is rejected, or until all the active treatments have had at least one stage. Therefore, in this calculation only the stages after every treatment has been added to the trial need to be considered, so $s^\star$ is defined as $s^\star=\max(S)$.  Using this gives for $j_0>s^\star$,
\begin{equation*}
\Psi_{j_0}= \int_{-\infty}^{\infty} \hdots \int_{-\infty}^{\infty} \Bigg{[} \prod_{k=1}^K \Bigg{(} \sum_{j=1}^{\min[J_k,j_0-s(k)]} \Phi(L_{k,j}(\theta_k),U_{k,j}(\theta_k),\Sigma_{k,j}) \Bigg{)} \Bigg{]}  d\Phi(t_1) \hdots d\Phi(t_{j_0}),
\end{equation*}
and for $j_0 \leq s^\star$  gives $\Psi_{j_0}=0$ and $\Psi_{0}=0$.
\item The probability the control treatment finishes  at each stage as the trial is stopped, given that a null hypothesis is rejected.
This is calculated by taking the difference between the probability that at least one null hypothesis is rejected by the control treatments ${j_0}$th stage, denoted by $\Upsilon_{j_0}$, and that at least one null hypothesis is rejected by the control's ${j_0}-1$ stage.
\begin{align*}
\Upsilon_{j_0}= 1- & \int_{-\infty}^{\infty} \hdots \int_{-\infty}^{\infty}   \prod_{k=1}^K \Bigg{[} \mathbbm{1}\{ j_0-s(k)>0 \} \bigg{[} \mathbbm{1}\{j_0-s(k)>1\} 
\\
& \;  \sum_{j=1}^{\min( J_k,j_0-s(k)-1 )}
 \Phi(L_{k,j}(\theta_k),U_{k,j}(\theta_k),\Sigma_{k,j}) + \mathbbm{1}\{ j_0-s(k)\leq J_k\}
\\
& \; \; \Phi(L_{h,j_0-s(k)}(\theta_k),\ddot{U}_{h,j_0-s(k)}(\theta_k),\Sigma_{h,j_0-s(k)}) \bigg{]} +  \mathbbm{1}\{ j_0-s(k)\leq 0\} \Bigg{]} d\Phi(t_1) \hdots d\Phi(t_{j_0}),
\end{align*}
where $\ddot{U}_{k,j}(\theta_k)= (u_{k,1}(\theta_k),\hdots u_{k,j-1}(\theta_k),u_{k,j}(\theta_k)).$ With $\Upsilon_{0}=0$.

\item The probability treatment $k'$ stops at each of its $J'$th stages because at least one other treatments null hypothesis has been rejected at this stage, denoted by $\Lambda_{k',J'}$ where,
\begin{align*}
\Lambda_{k',J'}=& \int_{-\infty}^{\infty} \hdots \int_{-\infty}^{\infty} \bigg{[} \varpi_{k',s(k')+J'-1}-\varpi_{k',s(k')+J'} \bigg{]} \bigg{[} \mathbbm{1}\{ J'-1>0\} 
\\
& \; \Phi(\ddot{L}_{k',J'-1}(\theta_k'),\ddot{U}_{k',J'-1}(\theta_k'),\Sigma_{k',J'-1}) + \mathbbm{1}\{ J'-1 \leq 0\} \bigg{]} d\Phi(t_1) \hdots d\Phi(t_{s(k')+J'}),
\end{align*}
where $\ddot{L}_{k,j}(\theta_k)= (l_{k,1}(\theta_k),\hdots l_{k,j-1}(\theta_k),l_{k,j}(\theta_k))$ and
\begin{align*}
\varpi_{k',j_0}= & \prod_{k=1, k \neq k'}^K \Bigg{[} \mathbbm{1}\{ j_0-s(k)>0 \} \bigg{[} \mathbbm{1}\{ j_0-s(k)>1\} \sum_{j=1}^{\min( J_k,j_0-s(k)-1 )}
 \Phi(L_{k,j}(\theta_k),U_{k,j}(\theta_k),\Sigma_{k,j}) +
\\
& \; \mathbbm{1}\{j_0-s(k)\leq J_k \} \Phi(L_{k,j_0-s(k)}(\theta_k),\ddot{U}_{k,j_0-s(k)}(\theta_k),\Sigma_{k,j_0-s(k)}) \bigg{]} + \mathbbm{1}\{ j_0-s(k)\leq 0 \} \Bigg{]},
\end{align*}
with $\varpi_{k',0}=1$.
\item The probability treatment $k'$ stops at each of its $J'$th stages because only $H_{0k'}$ is rejected, or no null hypotheses are dropped at this stage and treatment $k'$ is stopped as its test statistic drops below its lower boundary for that stage ($\Xi_{k',J'}$).
\begin{align*}
\Xi_{k',J'}= & \int_{-\infty}^{\infty} \hdots \int_{-\infty}^{\infty} \Bigg{(} \prod_{k=1,k \neq k'}^K \Bigg{[}\mathbbm{1}\{ s(k')+J'-s(k)>0 \} \bigg{(} \mathbbm{1}\{ s(k')+J'-s(k)>1 \}
\\
& \;  \bigg{[} \sum_{j=1}^{\min( J_k,s(k')+J'-s(k)-1 )}
 \Phi(L_{k,j}(\theta_k),U_{k,j}(\theta_k),\Sigma_{k,j})\bigg{]} + \mathbbm{1}\{ s(k')+J'-s(k)\leq J_k \}
\\
& \; \;   \Phi(L_{k,s(k')+J'-s(k)}(\theta_k),\ddot{U}_{h,s(k')+J'-s(k)}(\theta_k),\Sigma_{k,s(k')+J'-s(k)}) \bigg{)} + \mathbbm{1}\{ s(k')+J'-s(k)\leq 0 \} \Bigg{]} \Bigg{)}
\\
& \; \; \; \Bigg{[} \Phi(L_{k',J'}(\theta_{k'}),U_{k',J'}(\theta_{k'}),\Sigma_{k',J'}) + \Phi(L^{+}_{k',J'}(\theta_{k'}),U^{+}_{k',J'}(\theta_k'),\Sigma_{k',J'}) \bigg{]} d\Phi(t_1) \hdots d\Phi(t_{s(k')+J'}),
\end{align*}
where
\begin{align*}
U^{+}_{k,j}(\theta_{k'})=& (u_{k,1}(\theta_{k'}),\hdots u_{h,j-1}(\theta_{k'}), \infty)
\\
L^{+}_{k,j}(\theta_{k'})=& (l_{k,1}(\theta_{k'}),\hdots l_{h,j-1}(\theta_{k'}), u_{h,j}(\theta_{k'}))
\end{align*}
\end{enumerate}
Using the probabilities calculated above the expected sample size is:
\begin{align*}
N_E= \sum^W_{j_0=1}  (\Psi_{j_0} + \Upsilon_{j_0} - \Psi_{j_0-1} - \Upsilon_{j_0-1})n_{0,j_0} +\sum^K_{k'=1} \sum^{J_{k'}}_{J'=1} (\Xi_{k',J'} + \Lambda_{k',J'}) n_{k',J'}.
\end{align*}
\section{Table of results for Setting 1 based on the motivating trial}
As done for Setting 2 in Table 1 of the main paper in Web Table \ref{tab:setting1comparison} the results of the different comparison approach given in Section 3.2 is shown. Now for the expected duration until the MAMS trial finishes it is now assumed that the trial does not begin until the beginning of the second stage for Setting 1. Therefore 152 patients have already been recruited which equals 7.2 months.

\begin{table}[H]
\centering
 \caption{The results of the triangular stopping boundary shape on the design configuration for the motivating FLAIR trial under Setting 1 and three competing approaches: when running each trial separately, when using the MAMS design by \citep{MagirrD.2012AgDt} and when using the naive MAMS approach}
\begin{tabular}{c|c|c|c|c|c|c|c|c}

& FWER & LFC$_1$ &   NS$_1$  & $n_1$ & $\max(N)$ & $E(N|H_{G})$ & $E(N|\text{LFC}_1)$ & $E(N|\text{LFC}_2)$ \\
&  & LFC$_2$ &  NS$_2$  &  $n_2$ & $\max(T)$ & $E(T|H_{G})$ & $E(T|\text{LFC}_1)$ & $E(T|\text{LFC}_2)$ \\
\hline

Setting 1 & 0.025 & 0.802  & 2 & 76   & 540  & 351.8  & 285.8  &  400.8  \\
&  & 0.804 & 2   & 78  &  (25.7) &  (16.8) &  (13.6) &   (19.1) \\
\hline
Separate trials   & 0.025 & 0.805 & 2   &  77  & 616  & 368.2 & 419.3 & 419.3  \\
FWER control & & 0.805  & 2  &  77 &  (29.3) &  (17.5) &  (20.0) &  (20.0)  \\

\hline
Separate trials & 0.049 & 0.803 & 2   & 65 & 520  & 316.2  & 349.7  & 349.7  \\

no FWER control & & 0.803 & 2  & 65 &  (24.8) &  (15.1) &  (16.7) &  (16.7) \\

\hline
MAMS trial &0.025 & 0.804 & 2   & 76  & 456  & 280.7  & 309.8  & 309.8\\
2 Stage & & 0.804 & 2  & 76 & (29.0) &  (20.6) &  (22.0) &  (22.0) \\
\hline
Naive MAMS  & 0.048 & 0.802  & 3   & 65  & 455  & 299.3  & 234.0  & 331.4 \\
same $n_{j,k}$ &  &  0.787  &2   & 65 &  (21.7) & (14.3) & (11.1) &  (15.8) \\
\hline
Naive MAMS  & 0.047 & 0.676  & 2   &  65/26  & 260 & 197.8  & 172.8  & 214.4 \\
same $\max(N)$ & &  0.414 & 2   & 26 &  (12.4) &  (9.4) & (8.2) &  (10.2) \\
\end{tabular}
\vspace{0.5mm}
\\
\label{tab:setting1comparison}
Key: $E(N|H_{0.G})$, $E(N|\text{LFC}_k)$, $E(T|H_{0.G})$, $E(T|\text{LFC}_k)$ is the expected sample size and trial duration under the null and under the LFC for treatment $k$, respectively. 
\end{table}

\section{Tables of results based on the motivating trial for different stopping boundaries}
In Web Table \ref{tab:Scenario1} and Web Table \ref{tab:Scenario2} the results for the different combinations of stopping boundary shapes are shown for Setting 1 and 2 respectively. The stopping boundary shapes which are considered here are  Pocock \citep{PocockStuartJ.1977GSMi}, O'Brien and Fleming \citep{OBrienPeterC.1979AMTP} and Triangular stopping boundaries \citep{WhiteheadJ.1997TDaA}. However for both the Pocock boundary shape and the O'brien and Flemming boundary shape the symmetric futility boundary may be too stringent of a requirement to be able to drop ineffective treatments, therefore a simple alternative is $l_{k,j}=0$ for $j<J_k$ is used. As can be seen in these tables the upper and lower stopping boundaries are given, with the top row being the boundaries for the first active treatment added and the second row being for the second active treatment. 
\begin{landscape}    
\begin{table}[H]
\centering
\caption{The results of different stopping boundary shapes on the design configuration for the example trial under Setting 1.}
\begin{tabular}{c|c|c|c|c|c|c|c|c|c|c}
$\text{SB}_1$ & FWER & LFC$_1$ &   NS$_1$ & $L$ & $U$ & $n_1$ & $\max(N)$ & $E(N|H_{G})$ & $E(N|\text{LFC}_1)$ & $E(N|\text{LFC}_2)$ \\
$\text{SB}_2$ &  & LFC$_2$ &  NS$_2$ & & &  $n_2$ & $\max(T)$ & $E(T|H_{G})$ & $E(T|\text{LFC}_1)$ & $E(T|\text{LFC}_2)$ \\
\hline
OBF  & 0.025 & 0.803 &2 & \multirow{ 2}{*}{$\begin{pmatrix}
0 & 2.238 \\ 
0 & 2.238 \\ 
\end{pmatrix}$} &  \multirow{ 2}{*}{$\begin{pmatrix}
3.165 & 2.238 \\ 
3.165 & 2.238 \\ 
\end{pmatrix}$} & 69  & 491 & 383.7 & 322.9 &  430.6  \\
OBF  &  & 0.804 &2 & & & 71  & 23.4 & 18.3 & 15.4 &  20.5  \\
\hline
OBF  & 0.025 & 0.803 & 2  & \multirow{ 2}{*}{$\begin{pmatrix}
0 & 2.235 \\
0  & 2.435 \\ 
\end{pmatrix}$} & \multirow{ 2}{*}{$\begin{pmatrix}
3.160 & 2.235 \\
2.435  & 2.435 \\ 
\end{pmatrix}$} & 67  & 509 & 395.6 & 329.4 & 405.8 \\
 Po & & 0.805 & 2 & & & 77  & 24.2 & 18.8 & 15.7 & 19.3 \\
\hline
OBF  & 0.025 & 0.803 & 2 &\multirow{ 2}{*}{$\begin{pmatrix}
0 & 2.235 \\
0.832 & 2.354 \\ 
\end{pmatrix}$} & \multirow{ 2}{*}{$\begin{pmatrix}
3.161 & 2.235 \\
2.496 & 2.354 \\ 
\end{pmatrix}$} &  67  & 509 & 350.6 & 322.4 & 400.2 \\
 Tri & & 0.803 & 2 & & & 77  & 24.2 & 16.7 & 15.4 & 19.1 \\
\hline
Po & 0.025 &0.801 &2 &\multirow{ 2}{*}{$\begin{pmatrix}
0 & 2.442 \\
0 & 2.241 
\end{pmatrix}$} & \multirow{ 2}{*}{$\begin{pmatrix}
2.442 & 2.442 \\
3.169 & 2.241 
\end{pmatrix}$} & 76   & 514 & 403.1 & 285.5 &  448.1 \\
 OBF & & 0.805 & 2 & & & 70  & 24.5 & 19.2 & 13.6 &  21.3 \\
\hline
Po  & 0.025 & 0.804 & 2  &\multirow{ 2}{*}{$\begin{pmatrix}
0 & 2.440 \\
0 & 2.440
\end{pmatrix}$} & \multirow{ 2}{*}{$\begin{pmatrix}
2.440 & 2.440 \\
2.440 & 2.440 
\end{pmatrix}$} & 76   & 540 & 419.9 & 291.7 &  429.0 \\
Po & & 0.805 & 2 & & & 78  & 25.7 & 20.0 & 13.8 & 20.4 \\
\hline
Po & 0.025 &0.804 & 2 & \multirow{ 2}{*}{$\begin{pmatrix}
0 & 2.440 \\
0.834 & 2.358 
\end{pmatrix}$} & \multirow{ 2}{*}{$\begin{pmatrix}
2.440 & 2.440 \\
2.501 & 2.358 
\end{pmatrix}$} & 76   & 540 & 374.2 & 284.3 &  423.6 \\
 Tri & & 0.803 & 2 & &  & 78  & 25.7 & 17.8 & 13.5 &  20.2 \\
\hline
Tri  & 0.025 & 0.806 & 2 & \multirow{ 2}{*}{$\begin{pmatrix}
0.834 & 2.360 \\
 0 & 2.241	
\end{pmatrix}$} & \multirow{ 2}{*}{$\begin{pmatrix}
2.503 & 2.360 \\
3.170 & 2.241  
\end{pmatrix}$} & 77  & 515 & 381.9 & 287.8 &  425.9 \\
OBF &  & 0.802 & 2 & & & 69  & 24.5 & 18.2 & 13.7 &  20.3 \\
\hline
Tri   & 0.025 & 0.801 & 2 & \multirow{ 2}{*}{ $\begin{pmatrix}
0.834 & 2.358 \\
0 & 2.440
\end{pmatrix}$ } & \multirow{ 2}{*}{$\begin{pmatrix}
2.501 & 2.358 \\
2.440 & 2.440 
\end{pmatrix}$} & 76  & 536 & 394.5 &  292.2 &  403.9 \\
 Po &  & 0.800 & 2 & & &  77  & 25.5 & 18.8 &  13.9 &  19.2 \\
\hline
Tri  & 0.025 & 0.802 & 2  & \multirow{ 2}{*}{$\begin{pmatrix}
0.834 & 2.358 \\
0.834 & 2.358 
\end{pmatrix}$} & \multirow{ 2}{*}{$\begin{pmatrix}
2.501 & 2.358 \\
2.501 & 2.358 
\end{pmatrix}$} & 76 & 540 & 351.8 & 285.8 &  400.8 \\
Tri & & 0.804 & 2 & &  & 78  & 25.7 & 16.8 & 13.6 &  19.1 \\
\end{tabular}
\vspace{0.5mm}
\\
Key: $SB_k$ is the stopping boundary shape for treatment $k$; Tri is the triangular stopping boundary shape; OBF is the O'Brien and Fleming stopping boundary shape; Po is the Pocock stopping boundary shape; NS$_k$ is the maximum number of stages for active treatment $k$; $E(N|H_{0.G})$, $E(N|\text{LFC}_k)$, $E(T|H_{0.G})$, $E(T|\text{LFC}_k)$ is the expected sample size and trial duration under the null and under the LFC for treatment $k$, respectively.
\label{tab:Scenario1}
\end{table} 
\end{landscape}

\begin{landscape}
\begin{table}[H]
\centering
\caption{The results of different stopping boundary shapes on the design configuration for the example trial under Setting 2.}
\begin{tabular}{c|c|c|c|c|c|c|c|c|c|c}
$\text{SB}_1$ & FWER & LFC$_1$ &   NS$_1$ & $L$ & $U$ & $n_1$ & $\max(N)$ & $E(N|H_{G})$ & $E(N|\text{LFC}_1)$ & $E(N|\text{LFC}_2)$ \\
$\text{SB}_2$ &  & LFC$_2$ &  NS$_2$ & & &  $n_2$ & $\max(T)$ & $E(T|H_{G})$ & $E(T|\text{LFC}_1)$ & $E(T|\text{LFC}_2)$ \\
\hline
OBF  & 0.025 & 0.807 & 3 &\multirow{ 2}{*}{$\begin{pmatrix}
0 & 0 & 2.239 \\
0 & 2.231 &  - \\ 
\end{pmatrix}$}  & \multirow{ 2}{*}{$\begin{pmatrix}
3.878 & 2.742 & 2.239 \\
3.154 & 2.231 &  - \\ 
\end{pmatrix}$} & 41 & 440 & 334.3 & 333.6 & 367.0 \\
OBF  & & 0.800 & 2 & & & 69  & 21.0 & 15.9 & 15.9 & 17.5 \\
\hline
OBF  & 0.025 & 0.801 & 3 &\multirow{ 2}{*}{$\begin{pmatrix}
0 & 0 & 2.239 \\
0 & 2.433 & - \\ 
\end{pmatrix}$}   & \multirow{ 2}{*}{$\begin{pmatrix}
3.878 & 2.742 & 2.239 \\
2.433 & 2.433 &  - \\ 
\end{pmatrix}$} & 39 & 460 & 349.3 & 346.5 & 340.6 \\
 Po & & 0.802 & 2 & & &  76  & 21.9 & 16.6 & 16.5 & 16.2 \\
\hline
OBF  & 0.025 & 0.802 & 3 &\multirow{ 2}{*}{$\begin{pmatrix}
0 & 0 & 2.239 \\
0.831 & 2.351 & - \\ 
\end{pmatrix}$}   & \multirow{ 2}{*}{$\begin{pmatrix}
3.878 & 2.742 & 2.239 \\
2.494 & 2.351 & - \\ 
\end{pmatrix}$} & 39 & 460 & 313.5 &  332.9 & 336.3 \\
 Tri & & 0.801 & 2 & & & 76  & 21.9 & 14.9 &  15.9 & 16.0 \\
\hline
Po & 0.025 & 0.803 & 3  &\multirow{ 2}{*}{$\begin{pmatrix}
0 & 0 &2.547 \\
0 & 2.236 & - 
\end{pmatrix}$} & \multirow{ 2}{*}{$\begin{pmatrix}
2.547& 2.547& 2.547 \\
3.163 & 2.236 & - 
\end{pmatrix}$} & 48  & 476 & 358.8 & 298.4 &  391.2 \\
OBF  & & 0.805 & 2 & & & 71  & 22.7 & 17.1 & 14.2 &  18.6 \\
\hline
 Po & 0.025 & 0.806 & 3 &\multirow{ 2}{*}{$\begin{pmatrix}
0 & 0 & 2.547 \\
0 & 2.436  & - 
\end{pmatrix}$} & \multirow{ 2}{*}{$\begin{pmatrix}
2.547 & 2.547 & 2.547 \\
2.436 & 2.436  & - 
\end{pmatrix}$} & 47 & 496 & 373.4 & 308.4 &  362.8  \\
Po & & 0.802 & 2  & & & 77  & 23.6 & 17.8 & 14.7 &  17.3  \\
\hline
Po  &0.025 & 0.806 & 3 & \multirow{ 2}{*}{$\begin{pmatrix}
0 & 0 & 2.547 \\
0.832 & 2.355 & -
\end{pmatrix}$} & \multirow{ 2}{*}{$\begin{pmatrix}
2.547 & 2.547 & 2.547 \\
2.497 & 2.355 & -
\end{pmatrix}$} & 47  & 496 & 337.3 & 298.9 &  358.8  \\
 Tri & & 0.801 & 2 & & &  77  & 23.6 & 16.1 & 14.2 &  17.1  \\
\hline
Tri & 0.025 & 0.806 & 3 &\multirow{ 2}{*}{$\begin{pmatrix}
0 & 1.472 & 2.404 \\
0 & 2.235 &   -
\end{pmatrix}$} & \multirow{ 2}{*}{$\begin{pmatrix}
2.776 & 2.454 & 2.404 \\
3.161 & 2.235 &   -
\end{pmatrix}$} & 48 &  472 & 332.2 & 296.3 &  376.2  \\
OBF & & 0.801 &2 & & & 70  & 22.5 & 15.8 & 14.1 &  17.9  \\
\hline
Tri &0.025 &0.802 & 3  &\multirow{ 2}{*}{$\begin{pmatrix}
0 & 1.472 & 2.404 \\
0 & 2.435 &   -
\end{pmatrix}$} & \multirow{ 2}{*}{$\begin{pmatrix}
2.776 & 2.453 & 2.404 \\
2.435 & 2.435 & -
\end{pmatrix}$} & 46  & 492 & 347.0 & 306.9 &  353.2 \\ 
Po & & 0.804 & 2 & & & 77  & 23.4 & 16.5 & 14.6 &  16.8 \\ 
\hline
Tri & 0.025 & 0.802 & 3  &\multirow{ 2}{*}{$\begin{pmatrix}
0 & 1.472 & 2.404 \\
0.832 & 2.353 & - 
\end{pmatrix}$} & \multirow{ 2}{*}{$\begin{pmatrix}
2.776 & 2.453 & 2.404 \\
2.496 & 2.353 & -
\end{pmatrix}$} & 46 & 492 & 303.3 & 296.6 &  347.8  \\
Tri & &0.803 & 2 & & & 77  & 23.4 & 14.4 & 14.1 &  16.6  \\
\end{tabular}
\vspace{0.5mm}
\\
$SB_k$ is the stopping boundary shape for treatment $k$; Tri is the triangular stopping boundary shape; OBF is the O'Brien and Fleming stopping boundary shape; Po is the Pocock stopping boundary shape;  NS$_k$ is the maximum number of stages for active treatment $k$; $E(N|H_{0.G})$, $E(N|\text{LFC}_k)$, $E(T|H_{0.G})$, $E(T|\text{LFC}_k)$ is the expected sample size and trial duration under the null and under the LFC for treatment $k$, respectively.
\label{tab:Scenario2}
\end{table} 
\end{landscape}

\section{Distribution of sample size}
\begin{figure}[H]
\begin{subfigure}{1\textwidth}
  \centering
  \includegraphics[width=.9\linewidth,trim= 0 0.5cm 0 1cm, clip]{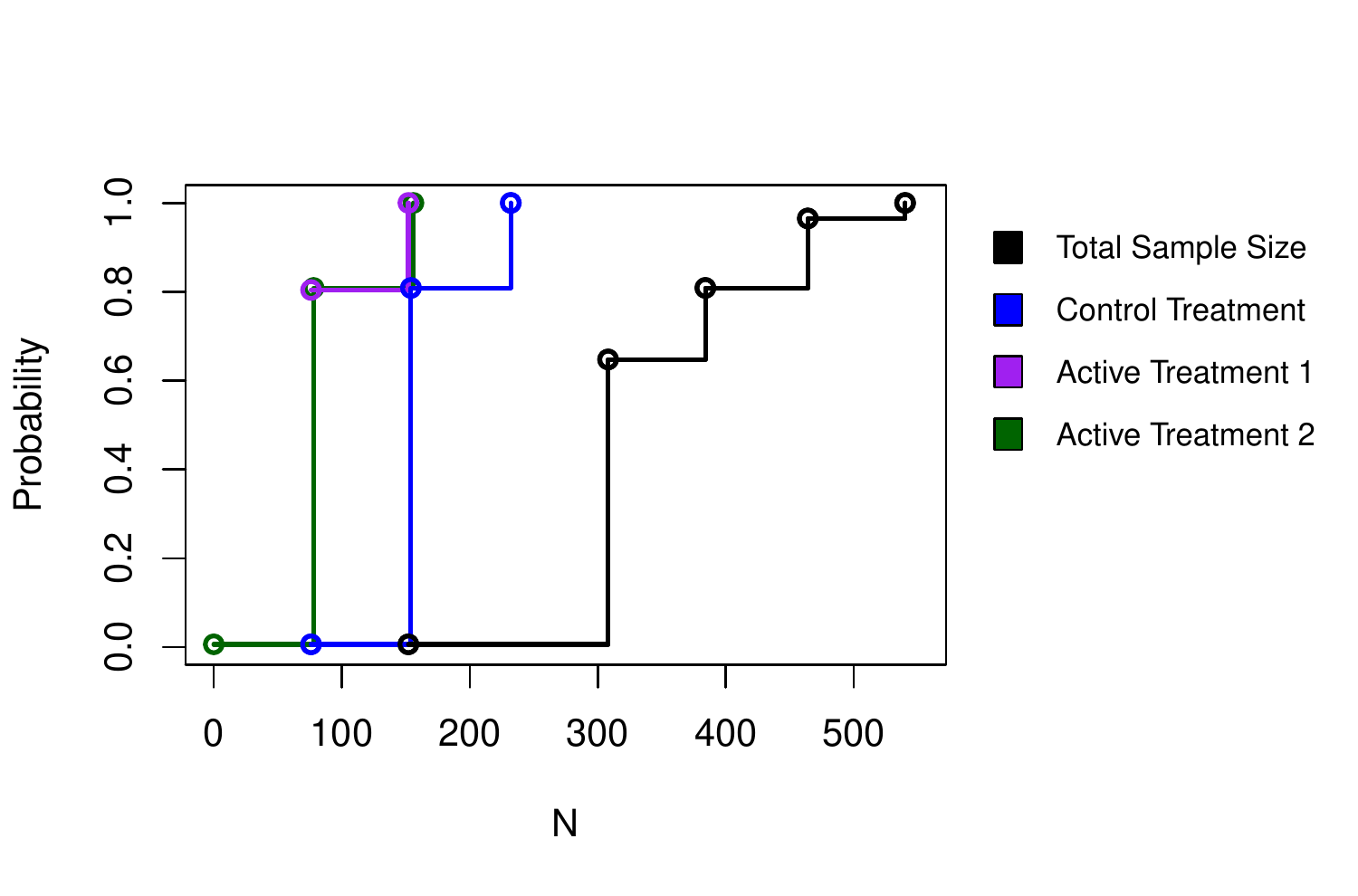}  
\end{subfigure}
\caption{Cumulative distribution functions (CDF) of the number of treatments needed for the trial in Setting 2 of the total sample size and for each arm individually}
\label{fig:2distribution}
\end{figure}

The distribution of the total sample size and the sample size of each treatment when the triangular stopping boundaries are used for Setting 1 under the global null is given in Web Figure \ref{fig:2distribution} with the probability mass function for the total sample size given in Equation \eqref{equ:PMF2}. In this example the triangular stopping boundaries for Setting 1 gives the interquartile range of 308 to 384 and median of 308 under the global null for the total sample size.  
\begin{equation}
\text{Distribution of the total sample size Setting 1} =
    \begin{cases}
     0.006, & \text{if}\ N=152 \\
     0.641, & \text{if}\ N=308 \\
     0.161, & \text{if}\ N=384 \\
     0.156, & \text{if}\ N=464 \\
     0.035, & \text{if}\ N=540 \\  
    \end{cases}
    \label{equ:PMF2}
\end{equation}
The probability mass function for the total sample size for Setting 2 is given in Equation \eqref{equ:PMF3}
\begin{equation}
\text{Distribution of the total sample size Setting 2} =
    \begin{cases}
     0.003, & \text{if}\ N=92 \\
     0.402, & \text{if}\ N=246 \\
     0.369, & \text{if}\ N=292 \\
     0.098, & \text{if}\ N=400 \\
     0.034, & \text{if}\ N=415 \\ 
     0.071, & \text{if}\ N=446 \\
     0.023, & \text{if}\ N=492 \\ 
    \end{cases}
    \label{equ:PMF3}
\end{equation}

\section{Robustness to the timing of the actual adding for Setting 1}
\begin{figure}[H]
\begin{subfigure}{1\textwidth}
  \centering
  \includegraphics[width=0.7\linewidth,trim= 0 0cm 0 1cm, clip]{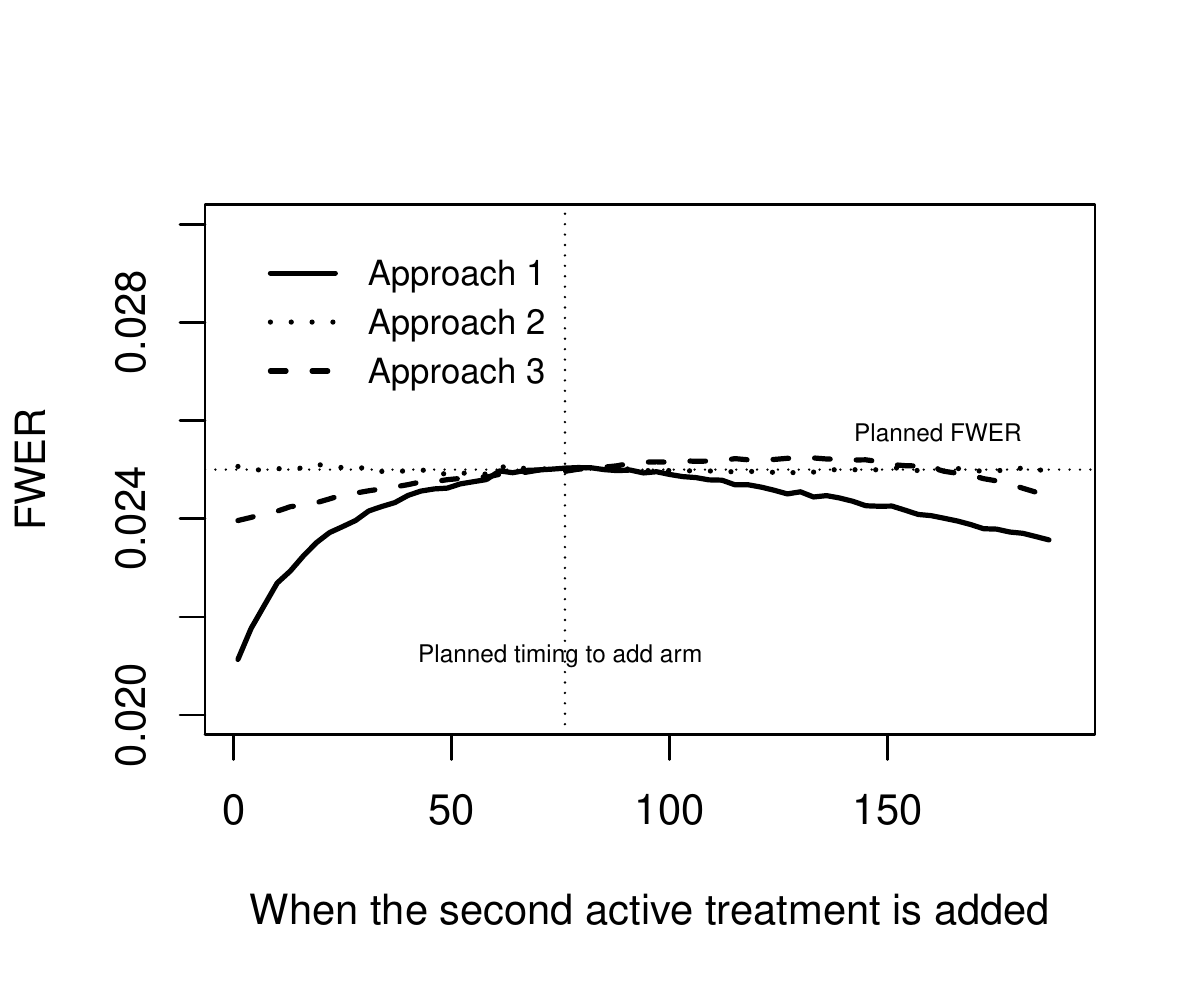}  
\end{subfigure}
\begin{subfigure}{1\textwidth}
  \centering
  \includegraphics[width=.7\linewidth,trim= 0 0cm 0 1cm, clip]{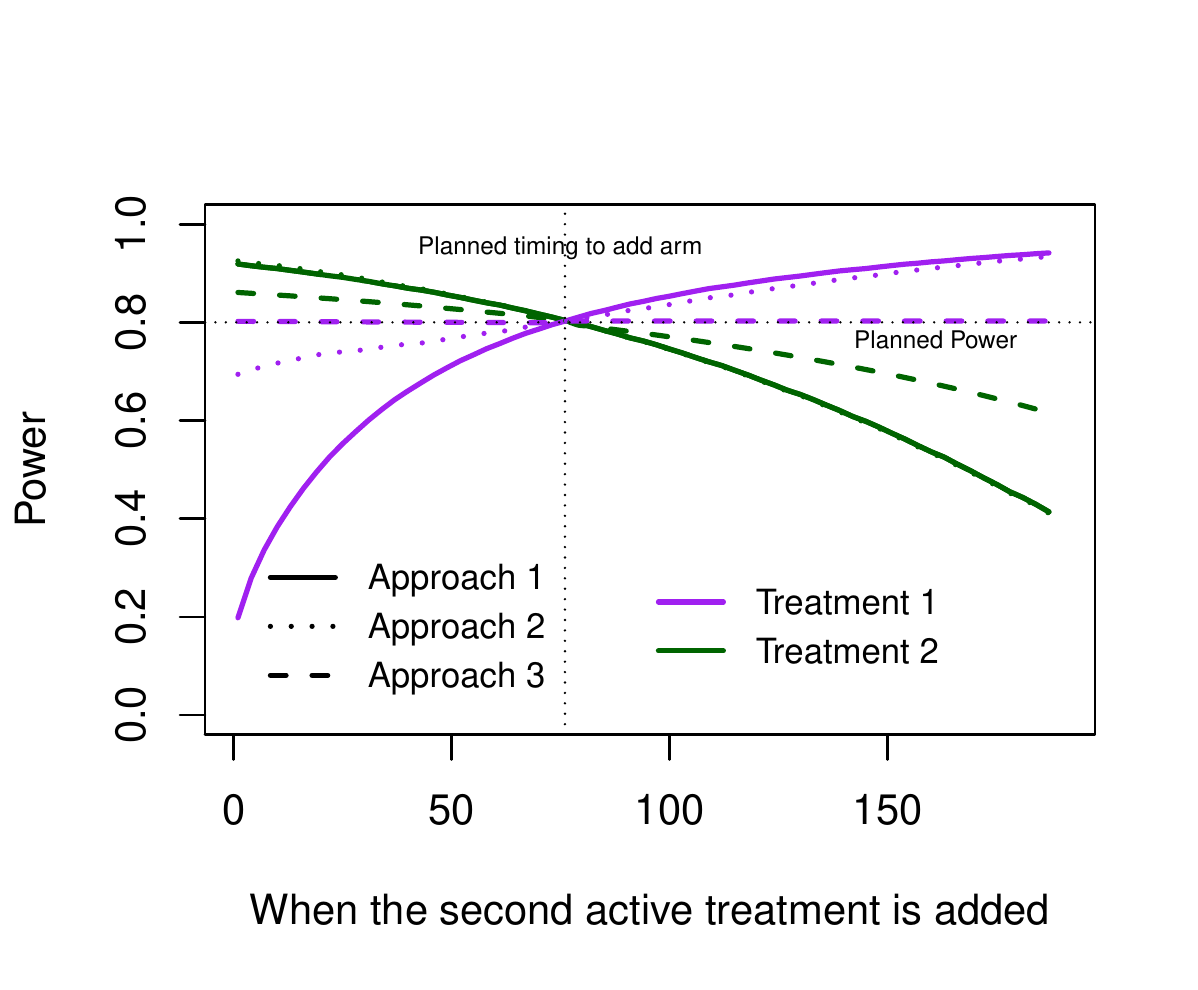}  
\end{subfigure}
\caption{The effect of adding the treatment later or earlier than planned using three different approaches for Setting 1 on power and FWER. With sub-figure (a) showing the FWER under the global null, and in sub-figure (b) showing the power under the LFC for Treatment 1 and the power under the LFC for Treatment 2.}
\label{fig:2stagesim}
\end{figure}

The same 3 approaches, as given in Section 4, to adding the second treatment earlier or later are studied here for Setting 1 as can be seen in Web Figure \ref{fig:2stagesim}. 

\section{The effect of a larger $\theta_0$ when using triangular stopping boundaries}
\begin{figure}[]
\begin{subfigure}{1\textwidth}
  \centering
  \includegraphics[width=.7\linewidth,trim= 0 0cm 0 1cm, clip]{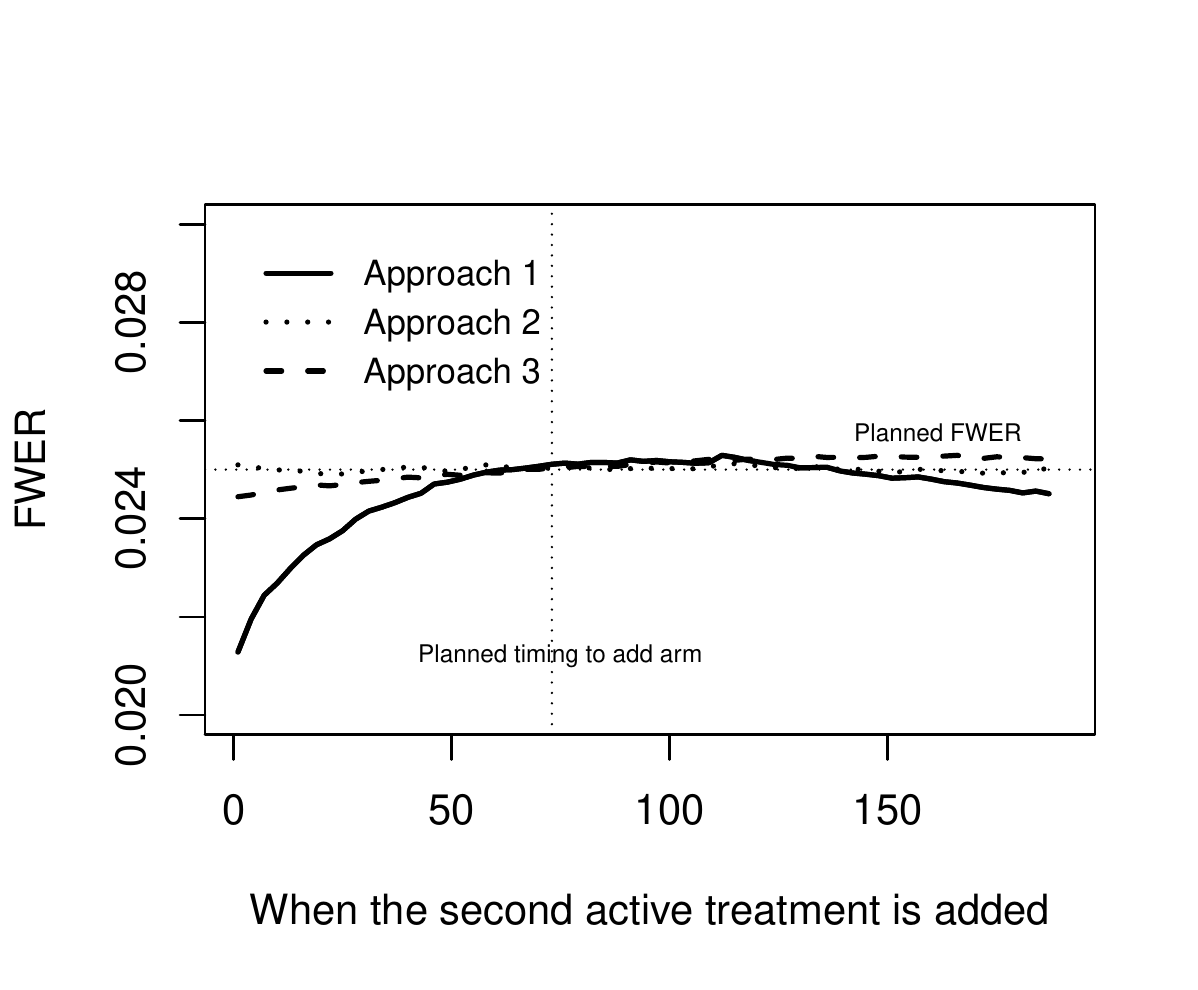}  
\end{subfigure}
\begin{subfigure}{1\textwidth}
  \centering
  \includegraphics[width=.7\linewidth,trim= 0 0cm 0 1cm, clip]{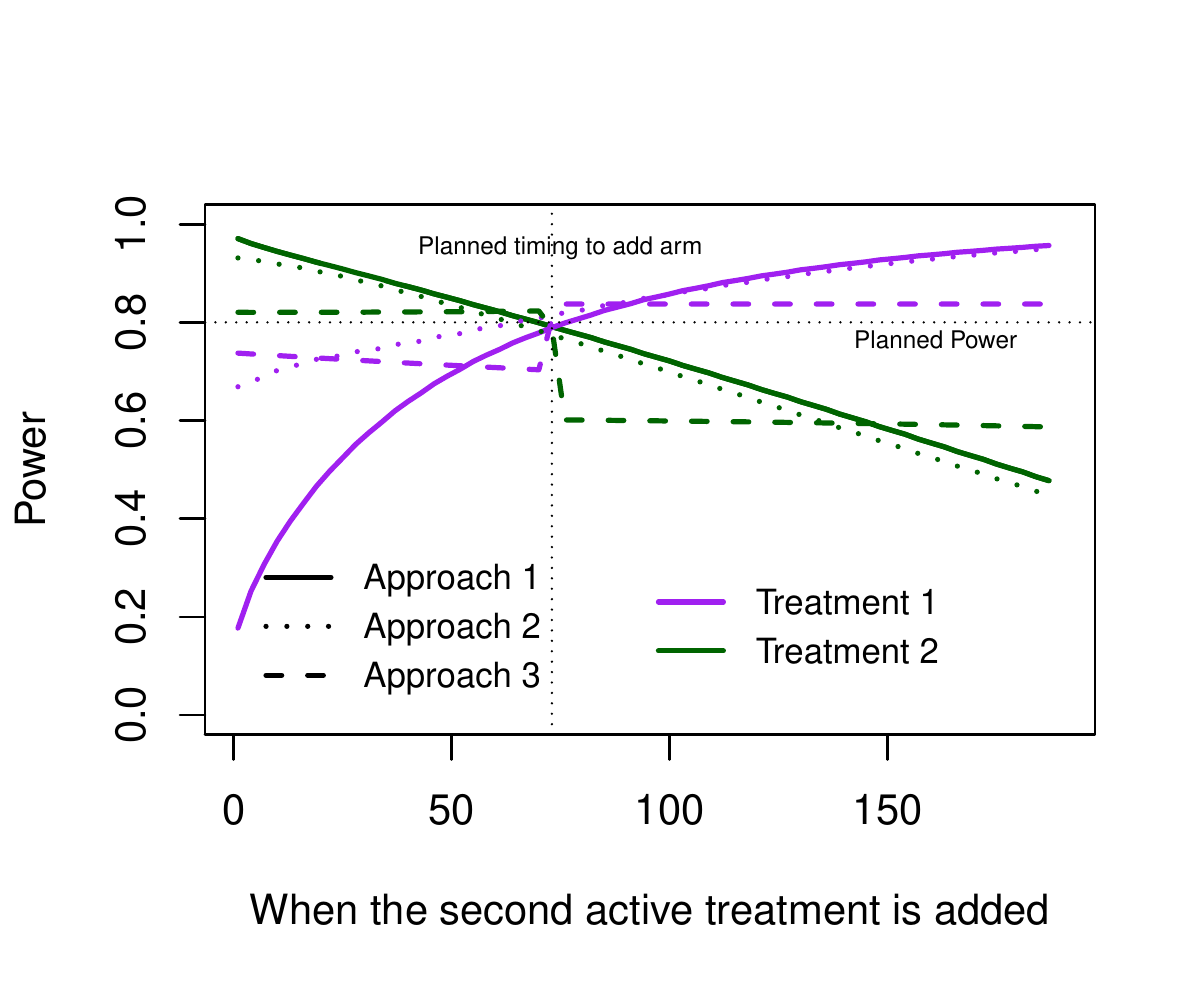}  
\end{subfigure}
\caption{The effect of adding the treatment later or earlier than planned using three different approaches for Setting 1 on power and FWER when $\theta_0=-\log(0.80)$. With sub-figure (a) showing the FWER under the global null, and in sub-figure (b) showing the power under the LFC for Treatment 1 and the power under the LFC for Treatment 2.}
\label{fig:2stagesimEx}
\end{figure}

\begin{figure}[]
\begin{subfigure}{1\textwidth}
  \centering
  \includegraphics[width=.7\linewidth,trim= 0 0cm 0 1cm, clip]{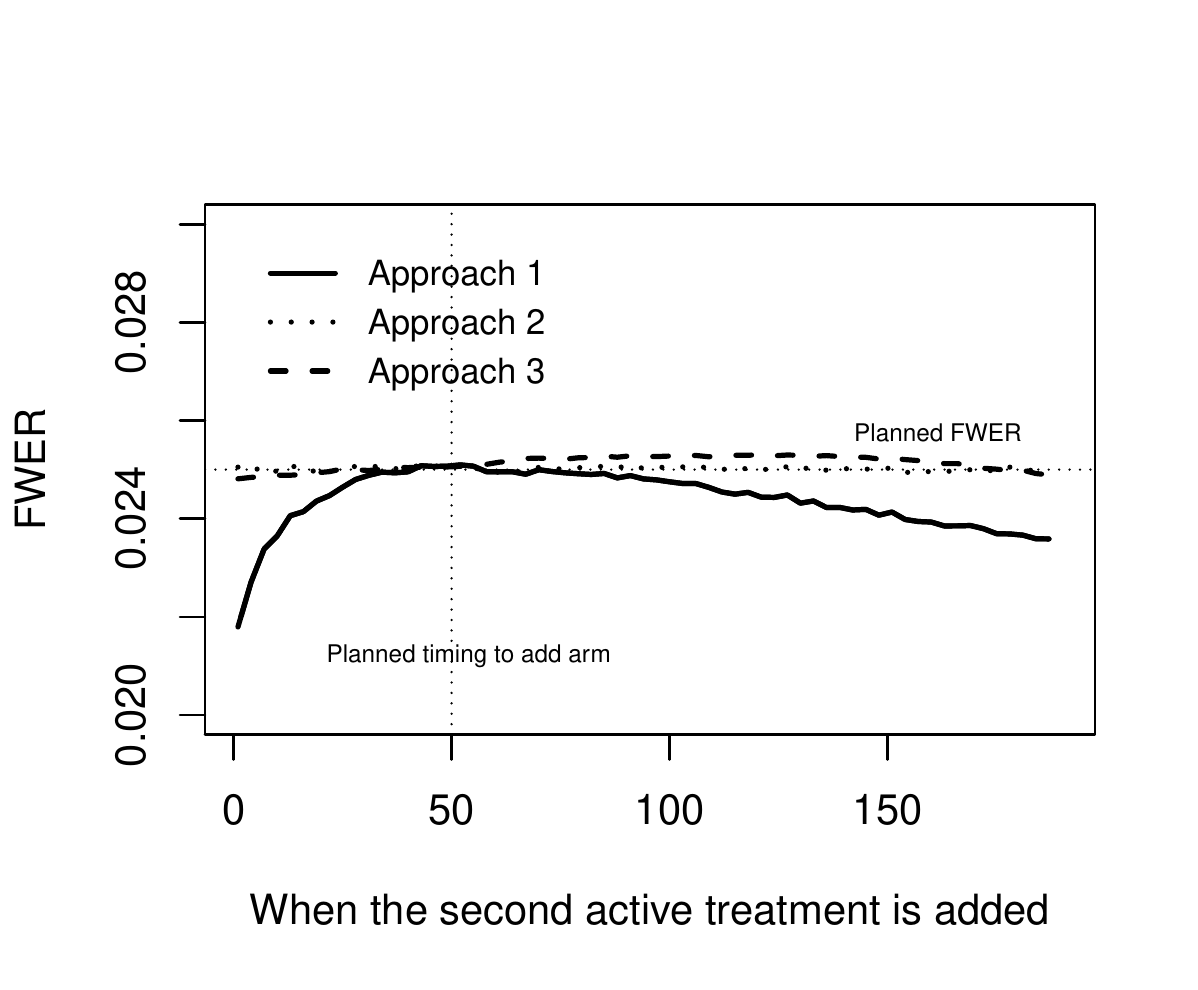}  
\end{subfigure}
\begin{subfigure}{1\textwidth}
  \centering
  \includegraphics[width=.7\linewidth,trim= 0 0cm 0 1cm, clip]{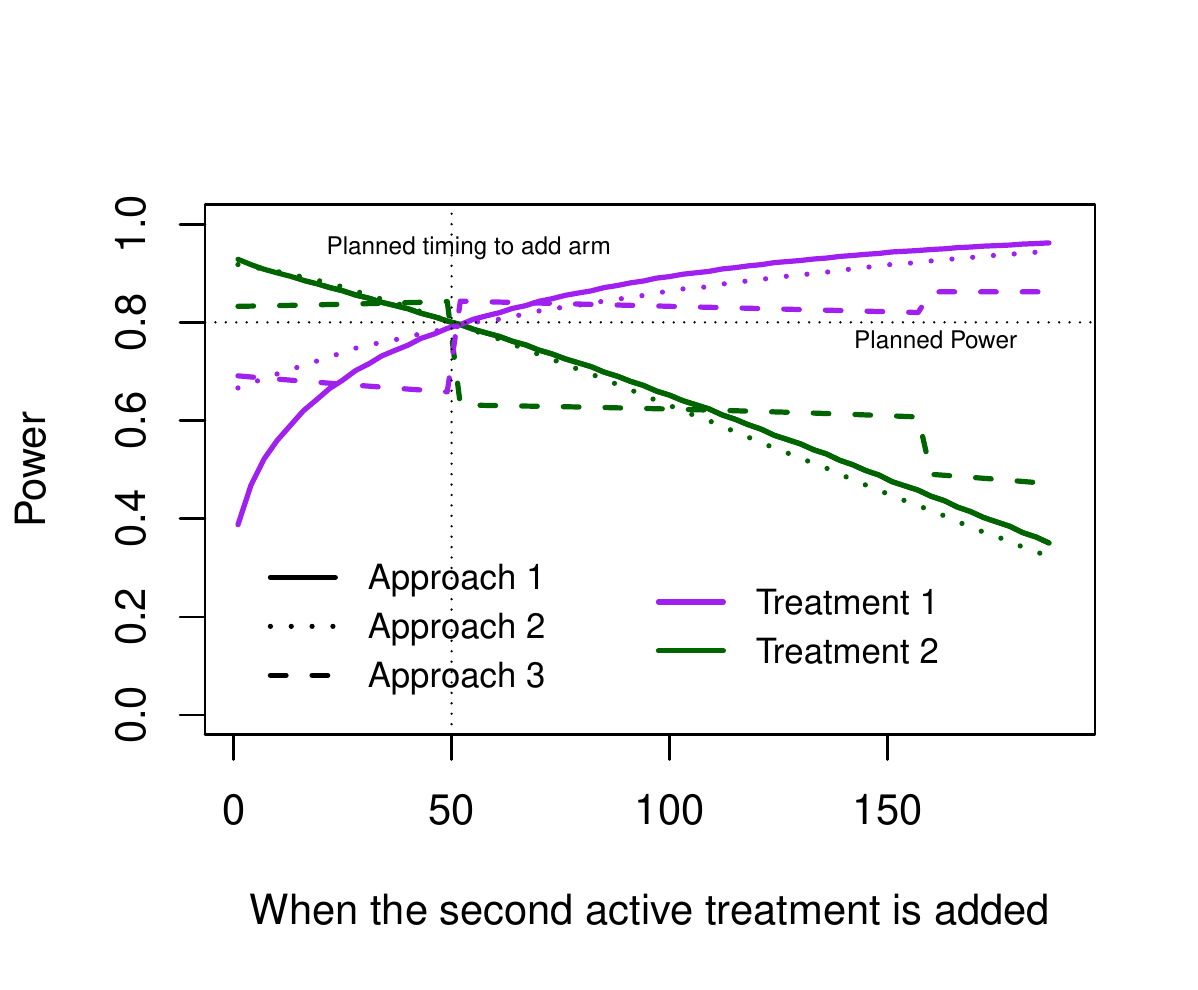}  
\end{subfigure}
\caption{The effect of adding the treatment later or earlier than planned using three different approaches for Setting 2 on power and FWER when $\theta_0=-\log(0.80)$. With sub-figure (a) showing the FWER under the global null, and in sub-figure (b) showing the power under the LFC for Treatment 1 and the power under the LFC for Treatment 2.}
\label{fig:3stagesimEx}
\end{figure}

For this example the same variables are used as the ones discussed in Section 3 apart from $\theta_0 = -\log(0.80)$. The main focus of this section is to show how this larger $\theta_0$ can result in the third approach to adding treatments earlier or later, as discussed in Section 4, does not control the power of all the treatments when the treatment is added earlier. The stopping boundaries, sample size and expected sample size for both settings are given in Web Table \ref{tab:supmat}. The effect of adding the treatment earlier or later than planned using the three approaches discussed in Section 4  is given in Web Figure \ref{fig:2stagesimEx} and Web Figure \ref{fig:3stagesimEx} for Settings 1 and 2, respectively. Both these figures now show that the power under the LFC for Treatment 1 is no longer controlled for any of the approaches when the second treatment is added earlier.   

\begin{table}[H]
\centering
\caption{The results of the triangular stopping boundaries on the design configuration for the example trial with $\theta_0=-\log(0.80)$ for both settings.}
\begin{tabular}{c|c|c|c|c|c|c}
Setting  & $n_1$ & $n_2$ & $\max(N)$ & $E(N|H_{G})$ & $E(N|\text{LFC}_1)$ & $E(N|\text{LFC}_2)$ \\
\hline
1 & 73 & 144  & 795 & 501.7 & 384.8 &  492.8  \\
\hline
2 & 50 &  108  & 632 & 388.3 & 374.8 &  422.7  \\
\end{tabular}
\vspace{0.5mm}
\\
Key: $E(N|H_{0.G})$, $E(N|\text{LFC}_k)$  is the expected sample size under the null and under the LFC for treatment $k$, respectively.
\label{tab:supmat}
\end{table} 

\par
\end{document}